%% file: Hunt_M81_source.tex
\documentclass[twocolumn]{aastex631}
\usepackage{soul}
\usepackage{enumerate}
\usepackage{graphicx}	
\usepackage{amssymb}	
\usepackage{multirow}   
\usepackage{array}      %
\usepackage{mathtools}
\usepackage{amsmath}
\makeatletter
\def\env@matrix{\hskip -\arraycolsep 
  \let\@ifnextchar\new@ifnextchar
  \array{*{\c@MaxMatrixCols}c}}
\makeatother

\usepackage{hanging, booktabs}
\usepackage[caption=false]{subfig}
\usepackage{times}
\usepackage{amsmath}
\usepackage{natbib} 
\usepackage{longtable}
\usepackage{afterpage}

\def\apjl{{ApJ}}                
\def\apjs{ {ApJS}}               
\def\ao{ {Appl.~Opt.}}           
             
\def\aap{ {A\&A}}

\def\arcsec{$^{\prime\prime}$}

\newcommand{\msun}{{$M_{\odot}$}}
\newcommand{\mstar}{{$M_{\star}$}}

\newcommand{\degree}{$^{\circ}$} 
\newcommand{\gtsima}{$\; \buildrel > \over \sim \;$}
\newcommand{\ltsima}{$\; \buildrel < \over \sim \;$}
\newcommand{\prosima}{$\; \buildrel \propto \over \sim \;$}
\newcommand{\gsim}{\lower.5ex\hbox{\gtsima}}
\newcommand{\lsim}{\lower.5ex\hbox{\ltsima}}
\newcommand{\simgt}{\lower.5ex\hbox{\gtsima}}
\newcommand{\simlt}{\lower.5ex\hbox{\ltsima}}
\newcommand{\simpr}{\lower.5ex\hbox{\prosima}}
\newcommand{\sqarcs}{arcsec$^{2}$}
\newcommand{\es}{erg~s$^{-1}$}

\newcommand{\lsun}{$L_{\odot}$}

\newcommand{\cxo}{\textit{Chandra}}
\newcommand{\hst}{\textit{HST}}

\newcommand{\lx}{$L_{\rm X}$}

\newcommand{\lm}{{LMXB}}
\newcommand{\hm}{{HMXB}}
\newcommand{\im}{{IMXB}}
\newcommand{\lms}{{LMXBs}}
\newcommand{\hms}{{HMXBs}}
\newcommand{\ims}{{IMXBs}}

\newcommand{\sn}{{S$_{N}$}}

\begin{document}
\title{Calibrating X-ray binary luminosity functions via optical reconnaissance II. The high-mass XLF and globular cluster population of X-ray binaries in the low star-forming spiral M81}
\author[0000-0002-4669-0209]{Qiana Hunt}
\affiliation{Department of Astronomy, University of Michigan, 1085 S University, Ann Arbor, MI 48109, USA}
\author{Elena Gallo}
\affiliation{Department of Astronomy, University of Michigan, 1085 S University, Ann Arbor, MI 48109, USA}
\author{Rupali Chandar}
\affiliation{Department of Physics and Astronomy, University of Toledo, Toledo, OH 43606, USA}
\author{Angus Mok}
\affiliation{OCAD University, Toronto, Ontario, M5T 1W1, Canada}
\author{Andrea Prestwich}
\affiliation{Harvard-Smithsonian Center for Astrophysics, 60 Garden Street, Cambridge, MA 02138, USA}

\begin{abstract}

We characterize the optical counterparts to the compact X-ray source population within the nearby spiral galaxy M81 using multi-band \textit{Hubble Space Telescope} (\hst) imaging data. By comparing the optical luminosities and colors measured for candidate donor stars and host clusters to stellar and cluster evolutionary models, respectively, we estimate the likely masses and upper age limits of the field and cluster X-ray binaries. We identify 15 low-mass X-ray binaries (i.e. donor star mass $\simlt$3 \msun) within ancient globular clusters, as well as 42 candidate high-mass X-ray binaries (i.e. donor star mass $\simgt$8 \msun). To estimate the likelihood of misclassifications, we inject 4,000 artificial sources into the \hst\ mosaic image and conclude that our classifications of globular clusters and high-mass X-ray binaries are reliable at the $>90$\% level. We find that globular clusters that host X-ray binaries are on average more massive and more compact than globular clusters that do not. However, there is no apparent correlation between the X-ray brightness of the clusters and their masses or densities, nor are X-ray binary hosts more X-ray luminous than the general field population of low-mass X-ray binaries. This work represents one of the first in-depth analyses of the population of X-ray binaries within globular clusters in a spiral galaxy. 

\end{abstract}

\section{Introduction} \label{sec:intro}
In the absence of an active galactic nucleus (AGN), the X-ray emission of galaxies is dominated by X-ray binaries (XRBs), with only minor contributions from coronally active binaries, cataclysmic variables, unresolved sources and supernova remnants (\citealt{fabbiano06,boroson11}, and references therein). Early population studies of high-mass XRBs (\hms, here defined as having donor masses in excess of $\gsim 8~M_{\odot}$) in star-forming galaxies established quantitative scaling rations between the host galaxy star formation rate (SFR) and the integrated X-ray luminosity of \hms\ \citep{grimm03,ranalli03,gilfanov-sfr}).  Along the same lines, \cite{gilfanov04} demonstrated that both the total number of low-mass XRBs (LMXBs, with donor masses below $\lsim 3~M_{\odot}$) and their integrated X-ray luminosity are proportional to the stellar mass (\mstar) budget of the host galaxy, thereby establishing a universal X-ray luminosity function (XLF) for \lms\ (see also \citealt{kim04}). 

Since these pioneering works, the \textit{Chandra X-ray Observatory} has collected several Msec of sub-arcsec resolution X-ray imaging data for tens of nearby galaxies. Updated XRB XLFs and their scaling relations are now routinely employed for a variety of purposes under the assumption that each population's XLF has a universal shape. For example, the total X-ray luminosity is taken as a reliable SFR proxy in distant, star-forming galaxies \citep{mineo14}, and the expected total X-ray luminosity in XRBs can be used to argue in favor or against a low-luminosity AGN on a statistical basis \citep{lehmer10,gallo10}. 

Recently, \cite{lehmer19} developed a global model to fit
simultaneously for the contributions from HMXBs, LMXBs, and
background X-ray sources using sub-galactic stellar-mass and star formation maps. This approach, applied to 38 nearby galaxies of varying morphological types and SFRs reveals a smooth decline in the XLF normalization per unit SFR, along with a decrease in
normalization at the high-luminosity end with increasing specific SFR (sSFR, defined as the ratio SFR/\mstar).
Further, \cite{lehmer19} note a possible discrepancy between the LMXB
XLF slopes below $\log L_{\rm X}\simlt 38$ and above $\log L_{\rm X}\simgt 39$ compared to the results obtained for early type ellipticals \citep{zhang12}. 
This is particularly interesting in light of the fact that, prior to \cite{lehmer19}, virtually all \hm\ XLF studies focused on galaxy samples with relatively high sSFR, and, as a result, treated all the detected compact X-ray sources outside of the bulge as \hms\ (e.g., \citealt{grimm03, gilfanov-sfr, mineo12}). It follows that, even two decades after the launch of \cxo, our knowledge of LMXBs in actively star forming galaxies remains relatively limited. \\ 

LMXBs are thought to form hundreds of
times more efficiently in globular clusters (GCs) than in the Galactic field \citep{Clark75,Katz75}, making GCs of particular interest in the study of LMXBs. 
Observations of massive ellipticals with the \textit{Chandra X-ray Observatory} and the \textit{Hubble Space Telescope} (\textit{HST}), have shown that between 20\% to 70\% of bright LMXBs ($\log L_{\rm X}\simlt 37$) currently reside in GCs \citep{angelini01, kundu02, jordan04, kundu07, humphrey08, peacock16}, and that a near constant 4\% - 6.5\% of GCs in ellipticals host bright LMXBs.
Given these extreme efficiencies, it has been suggested that field LMXBs may have formed in GCs that subsequently dissolved \citep{Grindlay84}, or were dynamically ejected from their natal GCs \citep{Grindlay85G,Hut92,Kremer18}. Recent work by \cite{lehmer20} support this scenario, although they also find evidence for an LMXB XLF component that scales with \mstar, suggesting that a non-negligible fraction of the LMXB population in ellipticals form `in-situ.'

The role that GCs play in the production of LMXBs in {spiral galaxies} remains some 20~years behind our understanding in ellipticals. 
The key impediment to progress in exploring the link between X-ray sources and GC in spirals 
has been the lack of reliable GC catalogs in all but a handful of spirals. Owing to on-going star-formation, the complex morphologies of spirals make it extremely challenging to identify compact stellar clusters in general, let alone to separate ancient GCs from reddened young clusters \citep{chandar04}. The fraction of young HMXBs in late-type galaxies is also poorly known, though $\approx 15$\% of XRBs in M101 \citep{chandar20} and $\approx25$\% in the Antennae \citep{rangelov12} are found in clusters younger than a few 100~Myr.

There is tentative evidence that the GC specific frequency may affect the shape and normalization of the \lm\ XLF, since GC \lms\ and field \lms\ likely have different evolutionary paths. In that case, late-type spiral galaxies, which tend to have lower GC specific frequencies, may yield lower \lm\ XLF normalizations than in elliptical galaxies, \citep{jordan04,kim06,sivakoff07,kim13,peacock17,luan18}. \\

This Paper is a continuation of an ongoing investigation into the XRB populations of nearby, late-type galaxies that leverages the combined imaging power of \cxo~and \hst. We use multi-band \hst\ imaging data to directly identify candidate optical counterparts to compact X-ray sources identified by \cxo. We previously applied this technique to classify XRBs within M101 \citep{chandar20} and M83 \citep{hunt21} and found that within a distance of $\sim$10 Mpc \hst\ is able to detect stars down to $\sim 3~$\msun, thereby allowing us to identify candidate donor stars of \ims\ and \hms. The source-by-source nature of our procedure means that our classification does not rely on any assumptions about the relationship with the local environment, such as \mstar\ and or SFR.  On the other hand, it is bound to suffer from some degree of contamination, as the accretion process affects the evolution, color and overall brightness of the donor stars.\\

Here, we apply our methodology to M81 (NGC 3031), shown in Figure~1. M81 is a spiral galaxy at an inclination of i = 58\degree\ \citep{okamoto15} and a distance of $\sim$3.6 Mpc \citep{lehmer19, lomeli21}, the nearest of our three galaxies sampled thus far. Compared to the previous galaxies we investigated, M81 has a low SFR of 0.25~\msun~yr$^{-1}$ (as opposed to 1.07 and 2.48 \msun~yr$^{-1}$ for M101 and M83 respectively) and a relatively high GC specific frequency (\sn) of 1.1 (as opposed to 0.43 and 0.17 for M101 and M83 respectively).
Furthermore, M81 allows us to take a deeper look into intermediate-mass XRB (\ims; with donor masses between 3-8 $M_{\odot}$) and how they are distributed throughout the galaxy. 

We present an optically classified catalog of X-ray sources within M81. Each source is identified as either an XRB (and tentatively classified by donor star mass), a possible supernova remnant, a background galaxy, or a foreground star. We quantify the likelihood of chance contamination using the results of an artificial source simulation in~\S\ref{sec:chance}. We describe the results of our source-by-source classification with a special focusing on the XRB spatial distribution (\S\ref{sec:sourceclass}), the X-ray luminosity function of high-mass XRBs (which we argue suffers from little contamination, \S\ref{sec:hmxlf}), and the association between XRBs and GCs in comparison to elliptical galaxies (\S\ref{sec:gcs}).

\section{Observations} \label{sec:obs}

\subsection{X-ray Source Catalog}\label{sec:xsources}

Following \citet{hunt21}, we make use of the X-ray point source catalog constructed by \citet[][hereafter L19]{lehmer19}, in which deep \cxo\ imaging data is examined for 38 nearby galaxies. This study includes thorough estimates of the X-ray completeness limits for each galaxy | crucial for our analysis. The \cxo\ data were reduced following the methods detailed in \citet{lehmer17}: the analysis was restricted to imaging data within 5\arcmin\ of the nominal center position of each galaxy, ensuring a sharp point spread function across the field of view. The source detection and parameter extraction were performed within 0.5-7 keV, where ACIS is best calibrated, and fluxes were converted to between 0.5-8 keV for the purpose of comparing the XLF analysis to previous studies (see L19 for details). 

\begin{figure*}
        \centering
        \includegraphics[width=0.8\textwidth]{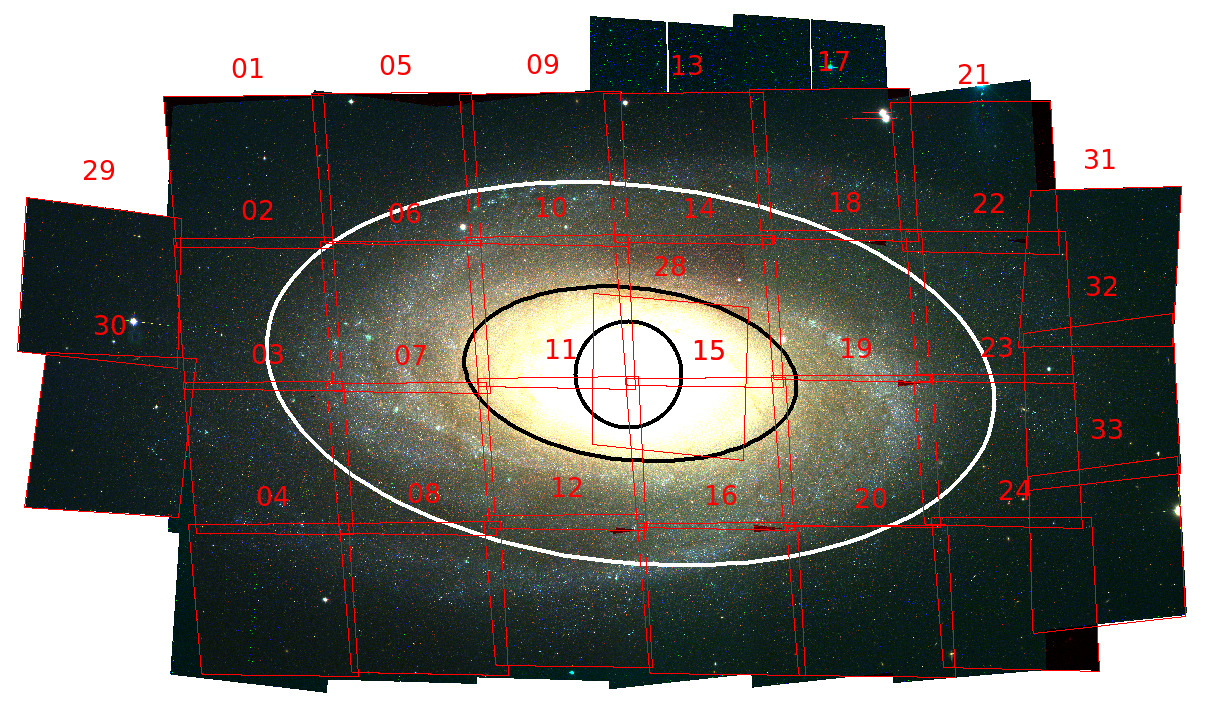}
        \caption{\hst\ ACS/WFC mosaic image of M81, composed of 33 individual fields numbered in red. The BVI color image is creating using F435W ($B$), F606W ($V$), and F814W ($I$) observations. The outer (white) ellipse traces the $K_{s} \approx 20$ mag arcsec$^{-2}$ galactic surface brightness contour \citep[see][]{jarrett03}. Within this region (which defines the chosen outer edge of the weak disk), L19 identify 199 compact X-ray sources, of which 150 are observed down to the (90\%) completeness limit of $\ell_X~\geq 36.3$. The inner (black) ellipse and circle represent the chosen edges of the bright disk and the bulge, respectively. \label{fig:mosaic}}
        \end{figure*}

Within M81, L19 identify 252 compact X-ray sources. The L19 study restricts its analysis of the subgalactic properties of the galaxy, and ultimately the XRB population models that arise from them, to the ellipse that traces the $K_{s}$=20 mag arcsec$^{-2}$ surface brightness contour outlined in white in Figure \ref{fig:mosaic} \citep[see also][]{jarrett03}. The ellipse (hereafter referred to as the L19 ellipse) has semi-major and semi-minor axes of 8.13 and 4.14 arcminutes respectively, covering a total area of $\sim$106 square arcminutes (383,409 square arcseconds) and includes 199 X-ray sources, of which 150 are observed down to the 90\% completeness limit of $\ell_X\geq 36.3$ (where $\ell_X$ represents logarithmic X-ray luminosities in units of \es). By comparison, the \hst\ footprint shown in Figure~\ref{fig:mosaic} includes 240 of the 252 compact X-ray sources from L19 (see \S\ref{sec:optdata} below). These include 41 sources that fall outside of the L19 ellipse, 36 of which are above the completeness limit. We will use the full 240 sample in our analysis unless otherwise specified.

\subsection{Optical Data}\label{sec:optdata}

Archival \hst\ imaging downloaded from the {Hubble Legacy Archive} (HLA\footnote{\texttt{http://hla.stsci.edu/}}) was used for the identification and classification of optical counterparts to the X-ray sources. For M81, a total of roughly 35 hours worth of ACS/WFC observations spanning 33 fields in three bands | F814W ($I$ band), F606W ($V$ band), and F435W for ($B$ band) | were obtained between September 2004 and March 2006 by J. Huchra (Prop. ID 10250) and A. Zezas (Prop. ID 10584). Of these, 27 fields have sufficient observations for this Work. Each field spans $3.4 \times 3.4$ arcminutes, yielding roughly 674,000~\sqarcs\ of coverage over the entire galaxy (for comparison, the L19 ellipse covers roughly 380,000~\sqarcs\ of the inner parts of the galaxy). 
The 33 fields were combined into a single mosaic using ASTRODRIZZLE from the STScI DrizzlePac software package\footnote{\texttt{https://www.stsci.edu/scientific-community/
software/drizzlepac.html}}. The full $BVI$ mosaic is shown in Figure \ref{fig:mosaic}. The mosaic was used to correct the astrometry between the \cxo\ and \hst\ data and to identify the candidate counterparts, while the individual fields, which preserved more reliable photometry, were used to measure source magnitudes.

A visual inspection of the \hst\ image reveals a bright central bulge and an outer disk. Unlike M83 and M101, the disk of M81 is marked by a region of exponential increase in brightness towards the bulge. As such, we find it useful to approach the X-ray source populations as members of 4 distinct regions | bulge, bright disk, weak disk, and outskirts | since the background brightness can drastically affect our ability to reliably classify the optical counterparts of X-ray sources (see \S\ref{sec:misclass}). The outer bounds of the weak disk is well-described by the L19 ellipse (white ellipse in Figure~\ref{fig:mosaic}). The edge of the bright disk is defined by the smaller black ellipse in Figure~\ref{fig:mosaic}, and has semi-major and minor axes radii of 3.70 and 1.91 arcminutes, respectively, so as to preserve the same aspect ratio as the L19 ellipse. The bulge (black circle in Figure~\ref{fig:mosaic}) is enclosed by a 1.18 arcminute radius circle as defined by \citet{fabricius12}.

\subsection{Identifying Candidate Optical Counterparts}\label{sec:optcount}

The first step to identifying candidate optical counterparts to X-ray sources is to correct the astrometry between the \cxo\ and \hst\ observations. 15 visually-identifiable background galaxies (i.e. those with clearly extended morphologies) and stellar clusters were selected as known reference points between the two data sets. We calculated a median relative positional offset between the X-ray coordinates of each source, as given by L19, and their \hst\ optical counterparts to obtain a median shift of 0.059 and -0.016 arcseconds along the \hst\ x- and y-axes, respectively, and standard deviations of 0.207 and 0.413 arcseconds. The X-ray centroid coordinates of all L19 sources in our sample were shifted by these offsets on the mosaic image. 

1- and 2-$\sigma$ uncertainty radii were calculated by adding in quadrature the standard deviation of the offsets with the positional uncertainties associated with the X-ray observation as given by Equations 14 and 12 in \citet{kim07}, corresponding to the 68\% and 95\%\ confidence radii. These equations take into account the total X-ray counts and the off-axis angle of the X-ray observations, which is important given the degradation of the point spread function of \cxo\ observations with distance from the pointing. The resulting 1- and 2-$\sigma$ radii, which are unique to each source, represent the regions within which we are most likely to detect an optical counterpart to a given X-ray source.

We then identify all potential optical counterparts within 2-$\sigma$ using the IRAF DAOFIND task to detect point-like sources on the composite images. The IRAF IMEXAMINE task is used to plot the radial profile of each source, which enables us categorize candidate counterparts as AGN, star clusters, or single stars (see \S\ref{sec:sourceclass}). To generate the magnitudes and colors needed for the analysis of each source, we perform aperture photometry on each detected source with IRAF PHOT. An aperture of 3 pixels is used on point sources, while  an aperture of 10 pixels is used for sources with radii much larger than 3 pixels (i.e. clusters). The photometry is performed on individual field images rather than on the mosaic, due to the brightness rescaling that occurs during the mosaic building process. The local background levels are determined in an annulus with radii between 20 and 25 pixels and subtracted from the photometric measurements. Aperture corrections of 0.357 ($V$), 0.329 ($B$), and 0.413 ($I$) were calculated as the median difference between the 3 and 20 pixel aperture magnitudes of several bright, isolated stars with smooth radial profiles that flatten towards the background sky magnitude. Similarly, the 10 pixel aperture corrections were 0.154 ($V$), 0.128 ($B$), and 0.160 ($I$). In both cases, an additional correction term of 0.941 (B), 0.947 (V), and 0.949 (I) accounting for the flux missing within a 20~pixel aperture was also added (see Encircled Energy Fractions from 20~pixels to infinity in \citealt{deustua17}). The magnitudes were converted to the VEGAMAG system by applying the zero-point magnitude for each filter as reported in Table 2 of \citet{deustua17}. Finally, the optical photometry was converted to absolute magnitudes at 3.63 Mpc, and we applied a blanket correction for foreground extinction equal to $A_{V}\approx0.255$~mag, assuming a galactic reddening of $E(B-V) = 0.08$ towards M81.
As for M83 \citep{hunt21}, we do not account for extinction intrinsic to each source, though we employ a confidence flag scheme (see Table \ref{tab:allsources} in the Appendix) to indicate sources that may be particularly susceptible to the effects of reddening and obscuration within M81.

\section{X-ray Source Classification}\label{sec:sourceclass}
 
X-ray sources may be classified as one of several objects on the basis of their detected optical counterparts: a foreground star, a background galaxy, a supernova remnant (SNR), or an XRB donor (either within a cluster or in the field). Foreground stars are identifiable by the bright diffraction spikes centered on the point source. Similarly, the majority of background galaxies that appear in the X-ray data can be identified by distinct morphological features, such as their extended radial profiles and the presence of a disk visible in the \hst\ image. For reference, 
at the typical depth of the X-ray observations that formed the basis for the L19 X-ray source catalog (X-ray fluxes $\sim10^{-15}-10^{-14}$ cgs), all optical counterparts of the X-ray sources in the sub-regions of the
Chandra Deep Field-S that were targeted with HST are found and
resolved at the resolution and depth corresponding to 1 orbit ACS
exposure \citep{grogin03,grogin05}. 

When identifying the most likely optical counterpart to X-ray sources, priority is given first to foreground stars and second to background galaxies that fall within the 2-$\sigma$ confidence regions of each X-ray source. These proximity-based classifications may be made with a high degree of confidence, as we find the probability of an unassociated chance superposition with either a foreground star or a background galaxy is rare within M81 (see \S\ref{sec:chance} for an in-depth analysis of chance superpositions with X-ray sources).  We identify 5 foreground stars and 10 background galaxies in our sample.
All other source classifications require a more in-depth analysis to ascertain. 

\subsection{Supernova Remnants}\label{sec:snr}
SNRs are a source of contamination that likely plagues many current XLFs. Recently, \cite{galiullin21} convincingly demonstrated that the vast majority of so-called soft and quasi-soft X-ray sources are indeed SNRs; these include soft X-ray sources both with and {without} an optical or radio counterpart SNR. As a confirmation, soft X-ray sources are a factor $\sim$~8 more abundant in star forming spiral galaxies compared to ellipticals.

In \citet{hunt21}, we devised a quantitative X-ray-based criterion by which to identify potential SNRs using the multi-wavelength SNR catalog compiled by \citet{long14} for M83 as the basis. We found that the majority of SNRs common between the \citet{long14} and L19 belong to a distinct parameter space on an \lx - X-ray hardness ratio (HR, here defined as the ratio of the sum of the counts in the 2-7 keV and 0.5-1.2 keV bands to the difference of their counts) plot. Based on this phenomenological approach, we adopt a minimum \lx\ and HR cuts to automatically select out candidate SNRs in our catalog, which correspond to $\ell_X \le 37.5$ and  HR$~\le-0.75$. 

We adopt the same \lx\ and HR cuts to identify potential SNRs in our M81 catalog, yielding 19 SNR candidates | roughly a factor 5 less than in M83. This is unsurprising considering the large difference in SFR between M81 and M83. We exclude an additional 6 SNRs identified by \citet{lee15}. Of the 25 total SNRs in our sample, 15 have optical counterparts, and 2 of these appear to be associated with clusters. 
 
        \begin{figure}[t]
            \centering
            \includegraphics[width=0.8\linewidth]{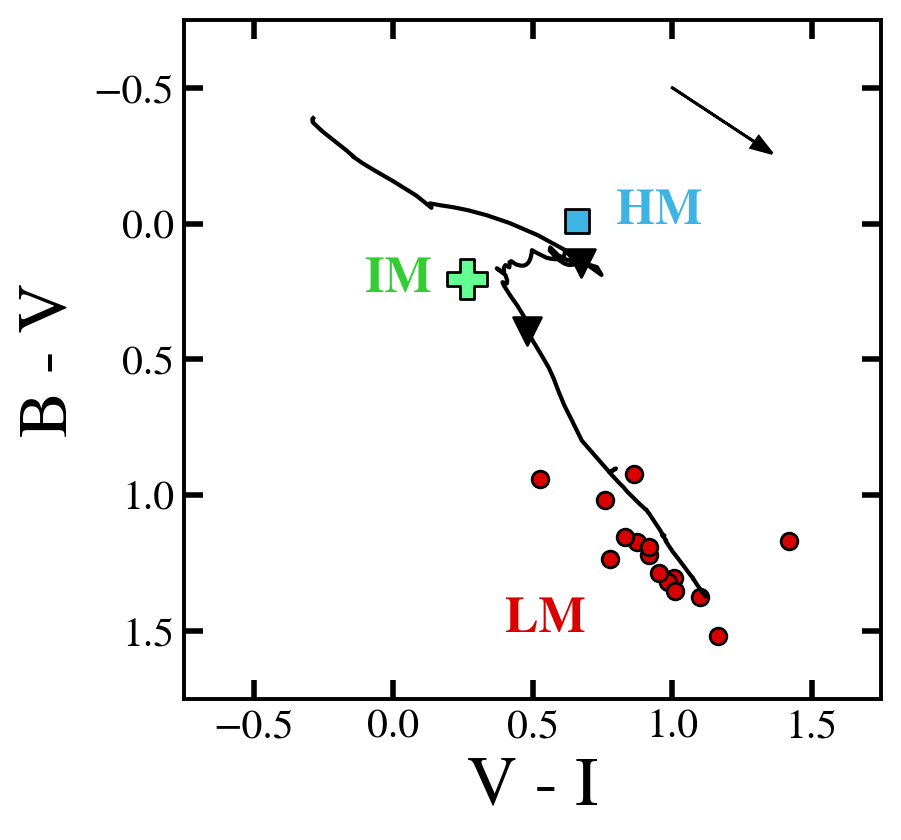}
            \figcaption{Measured $B-V$ vs $V-I$ colors of XRB host clusters compared with predictions for the color evolution of clusters from the \citet{bruzual03} models.  The black triangles mark cluster ages of 10 Myr (top) and 400 Myr (bottom).  The arrow represents the direction of reddening following the Milky Way extinction law with an example reddening of $E(B-V)=0.2$.   \label{fig:cluster}}
        \end{figure}

\subsection{Stellar Cluster Hosts}\label{sec:clusterages}

We expect to find some fraction of XRBs to exist within stellar clusters, due to the compact nature of clusters and increased likelihood of stellar interactions. These are identified as candidate sources with full-width half-maximums (FWHMs) that are broader than the point-spread function of a typical field star within the M81 \hst\ image; we find that GCs in M81 have FWHMs greater than 4 pixels (0.2 arcseconds), whereas typical stars have FWHMs less than 2 pixels (0.1 arcseconds). As a final confirmation, we compare our selections with a number of catalogs that identify clusters and extended sources in M81 \citep{nantais10, nantais11, nh10, sc10}, in which all but one of the sources we call a cluster are identified as clusters or extended sources. The only source that was not identified in any other catalog is found within the bulge region. 

We assume that, unless identified as SNRs by our method outlined in \S\ref{sec:snr}, X-ray emitting clusters are indeed XRB hosts. In order to classify a cluster XRB by donor mass, we use the cluster age as a proxy; since high-mass stars have hydrogen burning lifetimes of only $\sim10$~Myr and intermediate-mass stars have lifetimes of $\sim400$~Myr, globular clusters that are older than $400$~Myr are dominated by long-lived, low-mass stars, and their XRBs may be confidently classified as LMXBs. Likewise, we classify classify clusters younger than 10~Myr as \hm\ hosts and those falling between 10-400 Myr as \im\ hosts, since the most massive stars in a dense region are the most dynamically active and most likely to form stable binaries \citep{bonnell05}.

We approximate the ages of these clusters by comparing their colors to those predicted by the \citet{bruzual03} cluster evolution models at solar metallicity, as shown in Figure~\ref{fig:cluster}. The models span 1~Myr to 13~Gyr in cluster ages, moving from the upper left to the lower right. The black triangles demarcate the bounds of 10~Myr and 400~Myr, and the arrow shows the direction the colors would move due to reddening by dust. We find that 15 of the 17 clusters in our sample fall red-ward of the 400~Myr mark, distinguishing them as ancient globular clusters that host \lms. 1 cluster appears younger than 10~Myr, and 1 cluster falls between the two age ranges. We classify these as an \hm\ and an \im\ respectively. By comparison, M83 had 4 \hms, 4 \ims, and 4 LMXBs found in clusters, while M101 had 2 \hms, 6 \ims, and 1 \lms. The difference between these three galaxies is as expected given the SFR/\sn\ number of each.

\subsection{Field X-ray Binaries}\label{sec:donormass}

In the case where an X-ray source is neither a foreground star, background galaxy, SNR, nor a cluster, we assume the source is a field XRB with a donor star falling somewhere within the 2-$\sigma$ radius of the source. For sources that contain multiple candidates, the most likely donor is chosen on a case-by-case basis. Priority is given to high-mass stars that fall within or nearest to the 1-$\sigma$ radius, since high-mass stars are more likely to remain in a binary following a supernova kick \citep{kochanek19}. Low-mass stars are given second priority over intermediate-mass stars, since \ims\ are expected to be short-lived and are thus extremely rare \citep{pods02,pfhal02}. 

\begin{figure}
        \centering
        \includegraphics[width=0.45\textwidth]{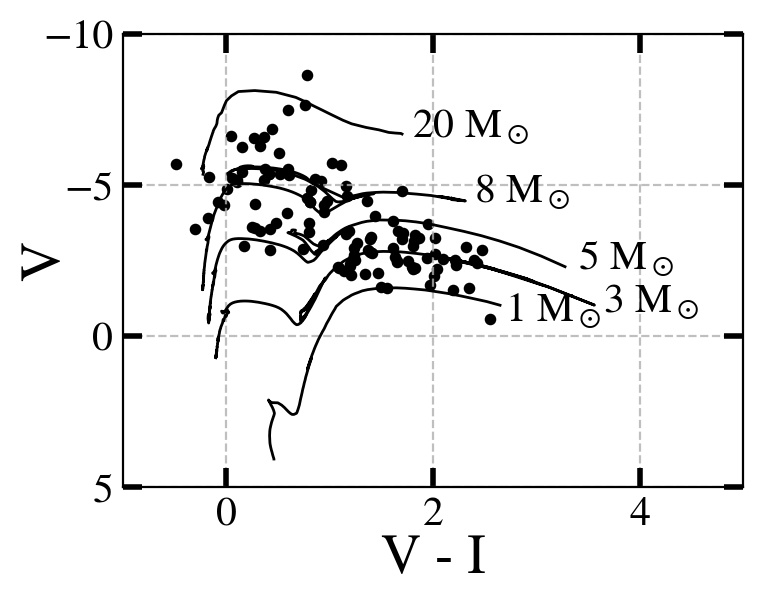}
        \caption{$V-I$ vs. $V$ color-magnitude diagram (CMD) of the donor stars of field XRBs in M81, compared to theoretical evolutionary tracks modeled at solar metallicity  \citep{bertelli94, girardi10, marigo17}. Donor stars in M81 are detectable with the \hst\ down to $\simgt 1$ \msun.
        \label{fig:CMD}}
        \end{figure}

To estimate the masses of candidate donor stars, we compare them against Padova theoretical stellar evolutionary models on a color-magnitude diagram (CMD, Figure \ref{fig:CMD}). Following our definitions of each category of XRBs and in the order of priority given, we classify sources with candidate donors above the 8~\msun\ model line as \hms, those with donors below the 3~\msun\ model line as \lms, and those falling between the two as \ims. The CMD demonstrates that at the distance of M81, the \hst\ image is deep enough to detect stars down to 1~\msun. Thus, sources that do not have visible donor stars are also categorized as \lms. This estimates implicitly assume no contamination from the X-ray-irradiated accretion disk to the optical band. This is further discussed in \S\ref{sec:misclass}. 

We note that M81 has regions of very intense star formation appearing as bright green HII regions in the \hst\ image. In these regions, it is difficult to detect individual stars due to crowding and saturation. In these cases, the donor star is assumed to be high-mass.

\subsection{Final Classifications}
Concluding the source classification, we identify 199 XRBs out of the total 240 X-ray sources that fall within the M81 \hst\ footprint: 122 \lms, 35 \ims, and 42 \hms. Of these, 17 exist within stellar clusters: 15 low-mass (within GCs), 1 intermediate-mass, and 1 high-mass. Taking only XRBs above the 90\% completeness limit of $\ell_X\geq 36.3$ estimated by L19, there are 159 XRBs: 100 \lms\ (11 in clusters), 25 \ims\ (none in clusters), and 34 \hms\ (1 in a cluster).
Of the remaining 41 sources, including the nucleus, 25 are SNRs, 10 are background galaxies, and 5 are foreground stars.

Restricting the sample to those sources that fall within the L19 ellipse, we find 88 \lms, 27 \ims, and 31 \hms in the field, with an additional 15 \lms, 1 \im, and 1 \hm\ within clusters. Above the completeness limit, this comes to  71 \lms, 19 \ims, and 26 \hms, with an additional 11 \lms\ in GCs. The L19 ellipse also contains 22 SNRs, 9 background galaxies, and 4 foreground stars (14, 5, and 4 above the completeness limit, respectively). 

A complete description of the positions, \lx, V-band magnitudes, optical colors, and classifications of all 240 sources within the \hst\ footprint is given in Table~\ref{tab:allsources}. Full-color optical images of all sources | with their 1- and 2-$\sigma$ radii outlined in red, candidate sources highlighted with yellow dashed circles, and most likely counterpart/donor circled in red | are shown in the Appendix, Figure~\ref{fig:mosaics}. 

The overall spatial distribution of the candidate XRB counterparts is shown in Figure~\ref{fig:sd}, where our classifications (not accounting for possible misclassifications; see \S\ref{sec:chance}) of low-, intermediate-, and high-mass donors are shown in red, green, and blue, respectively. The 15 cluster \lms\ are outlined in yellow.
As expected, the bulge of M81 is dominated by \lms: 39 total \lms\ (2 of which are found in GCs) are found in the bulge compared to 1 \hm. Above the completeness limit, there are 38 bulge \lms\ (1 in a cluster) compared to 1 \hm. The majority of \hms\ (31 out of 42) are within the disk; however, roughly $\sim52\%$ of all disk XRBs are \lms. When the disk is broken into a bright and weak component, 80\% of the bright disk XRBs are low-mass, while nearly 40\% of the weak disk XRBs are low-mass. The percentages may skew even higher, considering the possibility that some (or all) of the sources we classified as \ims\ may actually be \lms, as discussed in \S\ref{sec:chance}. 
 
When compared with the stellar mass, SFR and specific SFR (sSFR) sub-galactic maps generated by L19 (Figure~\ref{fig:lehmers}), \hms\ appear to be concentrated in high SFR regions, whereas most (albeit not all) \lms\ appear to follow the radial profile of the stellar mass. The \ims\ population seems to be somewhat hybrid; whereas their spatial distribution resembles more closely that of \hms, they are found in larger proportions in the bright disk. As discussed in \S\ref{sec:misclass}, however, this population is likely affected by a high degree of contamination, so we refrain from further speculating on their properties.

\section{Assessing Misclassifications}\label{sec:misclass} 

Here, we discuss a few of the most pressing sources of misclassification of the optical counterparts in M81. For an additional discussion of potential sources of misclassification inherent to our method, the reader is referred to \citet{chandar20} and \citet{hunt21}.

\subsection{Quantifying Chance Superpositions}\label{sec:chance}

In this section we assess the likelihood that an X-ray source may have a chance superposition (rather than physical association) with a high- or intermediate-mass star, globular cluster, or background galaxy. 
Since low mass stars dominate the stellar population of a galaxy, we seek to quantify how certain can we be that, for any given region, the true donor is not an unseen low-mass star. 
While our prioritization scheme for sources with multiple point-source candidates places \hms\ above other classifications in any case (due to the higher chance of binary survival post-supernova, \citealt{kochanek19}), this becomes a greater issue for \ims, which we would naturally de-prioritize compared to LMXBs due to their rareness. 

In order to quantify the frequency of chance superposition with a source that is \textit{not} a low mass star, we randomly populate the 4 regions of the galaxy (bulge, bright disk, weak disk, and outskirts, as defined in \S~\ref{sec:optdata}) with 1,000 artificial X-ray sources per region and classify each using the \hst\ images and photometry, as done with our `true' X-ray sources. 
The $2\sigma$ positional uncertainty used to search for an optical counterpart to each synthetic X-ray source is assumed to be the median of the $2\sigma$ positional uncertainties of the `true' X-ray sources within that region; these values are compiled in Table~\ref{tab:sim}. As expected, these increase as one moves further from the center of the galaxy, largely due to the degradation of the \cxo\ PSF farther from the pointing.

\begin{figure}[h]
    \centering
    \includegraphics[width=0.45\textwidth]{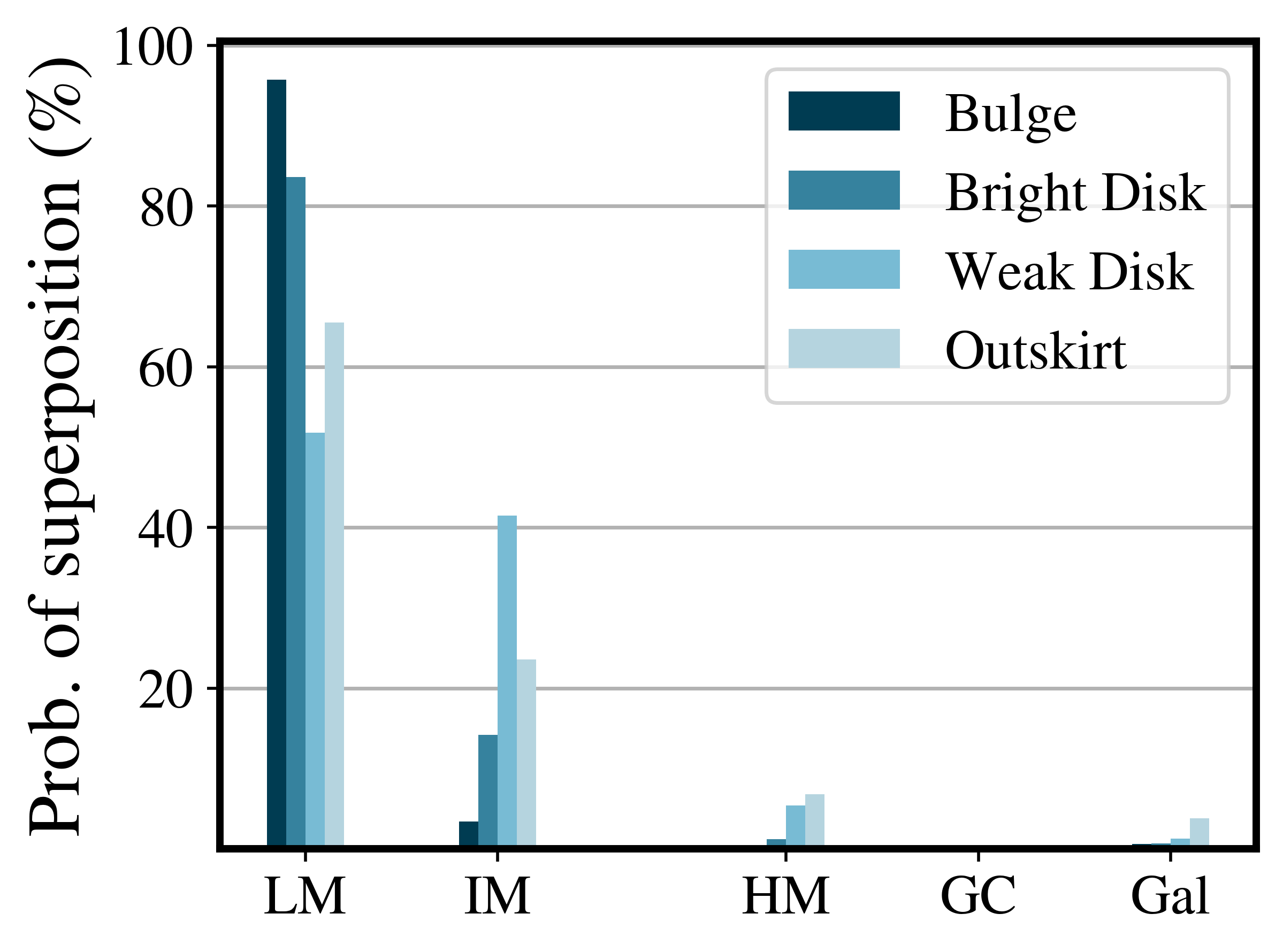}
    \caption{Results from our chance superposition simulation, in which 1,000 simulated sources were randomly distributed throughout each of our 4 regions | the bulge, the bright disk, the weak disk, and the outskirt, as defined in \S\ref{sec:optdata} | and their optical counterparts classified using the same methods as for the actual (`true') X-ray sources. These probabilities are also given in Table~\ref{tab:sim}.}
    \label{fig:sim_graph}
\end{figure}

The results of our artificial source experiment are compiled in Table~\ref{tab:sim} and shown in Figure~\ref{fig:sim_graph}. For each region we estimate the probability of a chance superposition with a high-mass star, an intermediate-mass star, a globular cluster, and a background galaxy as a percentage.

In Table~\ref{tab:pop}, we compile the number of observed XRBs that were classified as \hms, \ims, and \lms~ in the same four regions, as well as X-ray sources found in GCs and associated with background galaxies. The total number in each region is given as the first number in the table, and the number of sources above the completeness limit is given in parentheses.

One key result of the simulation is that the probability of finding a chance superposition with a high-mass star, a globular cluster, or a background galaxy is extremely low ($<1$\% in total) in the bulge region. There is a somewhat higher probability of a chance superposition with an intermediate-mass star ($3.4$\%). Because the bulge region of M81 is bright, it is hypothetically possible for sources to be misclassified as an \lm\ with no optical counterpart, as the background brightness could outshine any potential point source. However, the fact that we \textit{do} see a handful of intermediate-mass stars within our simulation in this region indicates that we are still able to identify stars down to intermediate masses, though intermediate-mass stars closer to the 3 \msun\ limit may be lost. This, as well as the low probability of chance superposition with non-low-mass sources, suggests that our source classifications | particularly of any \hms\ |  in the bulge are likely accurate and are not prohibitively sensitive to background brightness or chance superposition. 

The bright disk also shows a low probability for a chance superposition with a high-mass star (1.2\%), a globular cluster ($<1$\%), or a background galaxy ($<1$\%).  There is however, a more significant probability that an X-ray source will have a chance superposition with an un-associated intermediate-mass star ($14.2$\%). These probabilities increase in the weak disk, especially for high-mass ($5.4$\%) and intermediate-mass stars ($41.5$\%). The probability of overlapping with a background galaxy increased to 1.3\%, while there is virtually no chance superposition with a cluster in the weak disk. Potential explanations for these increases is the inclusion of the spiral arms in these regions, which tend to harbor bright, massive stars, the decreasing background brightness, and (perhaps more importantly) the larger positional uncertainty of the X-ray observations in this portion of the galaxy. 

The outskirts of the galaxy also show substantial contamination by intermediate-mass (23.6\%) and | albeit to a lesser degree | high-mass stars (6.8\%). The chance of overlapping with a background galaxies triples (3.8\%). The increased frequency of background galaxies in our simulation with galactocentric distance is not surprising, as many more background galaxies are visible in the outskirts of M81, where the background brightness is lowest and the density is most diffuse (see also \S\ref{sec:cxb}).

Taken together, these results lead to the following conclusions:
\begin{itemize}
\item The probability of finding a chance superposition with a globular cluster is negligible all 4 regions of M81. This indicates that any X-ray source-GC association is almost certainly physical. 

\item X-ray sources classified as \hms\ in the bulge (1~source) or bright disk (2 sources) are almost certain to be correctly classified. The weak disk and outskirts have relatively low probabilities of chance superposition with high-mass stars ($<10\%$ each). There are 70 and 31 XRBs above the completeness limit in the respective regions, with 23 and 7 identified as \hms\ in each region, respectively.
The artificial experiments suggest minor contamination, with at most 4 possible misclassifications in the weak disk and 3 in the outskirts.
\smallskip

We note that high-mass stars are more likely to survive in an X-ray binary than those with lower masses, so in a galaxy with continuous star formation, it may be reasonable to assume a greater fraction of XRBs identified as having high-mass donors are \textit{truly} \hms\ than this simple simulation suggests.  

\item At face value, intermediate-mass stars are, after low-mass stars, the most abundant potential counterparts within our simulation. This suggests a good number of our XRBs may overlap with intermediate-mass stars that are un-associated with the X-ray emission, simply by chance.
In the bright and weak disks, we classified 9 and 17 X-ray sources above the completeness limit as \ims\ (there are only a handful of such sources in the outskirts and none in the bulge). 
In the weak disk where the probability of chance superposition is greatest, we may expect up to 30 of the 70 XRBs to be misclassified as \ims. This suggests that we cannot trust our classified \im\ population to be truly representative of the \im\ population of M81. 

Evolutionary arguments further strengthen our conclusion that a large fraction of XRBs with an associated intermediate-mass stars may not be reliable. Even though the 
probability of \textit{forming} an XRB with an initial intermediate-mass donor is higher than that of a low-mass donor, population synthesis models show that most
intermediate-mass donors (those with neutron star accretor, that is) evolve very quickly into low-mass stars
through a short-lived thermal mass transfer phase \citep{pods02,pfhal02}.  This results into a population of
abnormally hot and luminous LMXBs | albeit with IMXB progenitors | which is bound to 
be misclassified through our method. 

We check whether X-ray irradiated disks may contribute significantly to our optical bands by comparing the colors of our \hms\ to other nearby, high-mass stars. We identify a sample of 115 high-mass comparison stars located within $\approx 3$ arcseconds of our 42 \hms, to account for potential environmental effects that could alter the colors of the donor stars. We find that the B-V and V-I colors of the candidate \hm\ donor stars are entirely consistent with typical field stars of similar brightness. We note, however, the discrepancies arising from, e.g. X-ray irradiated disks, would become more prominent in the UV band \citep{tao11}.

\item While remaining small, the probability of chance superposition with an optically identified background galaxy in our simulated sample increases radially towards the outskirts, as expected. In terms of our `true' sample, we find 3 X-ray sources associated with background galaxies in the bright disk and 6 in the weak disk, and only a single such object in the outskirts. This is addressed specifically in \S\ref{sec:cxb}.

\end{itemize}

\begin{table}[ht]
	\caption{Probability of Chance Superposition (percentage)}\label{tab:sim}
	\centering
	\begin{tabular}{lccccc}
	Region & Median & HMXB & IMXB & Globular & Background \\ 
	in M81 & $2\sigma$ & & & Cluster & galaxy \\ \hline \hline
Bulge & 0.\arcsec 54 & 0.1 & 3.4 & 0.1 & 0.6 \\
Bright disk & 0.\arcsec 66  & 1.2 & 14.2 & 0.2 & 0.7 \\ 
Weak disk & 0.\arcsec 78  & 5.4 & 41.5 & 0.0 & 1.3 \\
Outskirts & 1.\arcsec 48  & 6.8 & 23.6 & 0.1 & 3.8 \\
        \hline
	\end{tabular}
	\end{table}

\begin{table}[ht]
	\caption{Observed Populations (Total \& Completeness limited)}\label{tab:pop}
	\centering
	\begin{tabular}{lcccccc}
	Region & LMXB & HMXB & IMXB & Globular & Background \\ 
	in M81 & (Field) & & & Clusters & galaxy \\ \hline \hline
Bulge & 37 (32) & 1 (1) & 0 (0) & 2 (1) & 0 (0) \\
Bright disk & 25 (14) & 2 (2) & 6 (2) & 7 (5) & 3 (1) \\ 
Weak disk & 26 (25) & 29 (23) & 22 (17) & 6 (5) & 6 (4) \\
Outskirts & 19 (18) & 10 (7) & 7 (6) & 0 (0) & 1 (1) \\
        \hline
	\end{tabular}
	\end{table}

\begin{figure}
    \centering
    \includegraphics[width=0.4\textwidth]{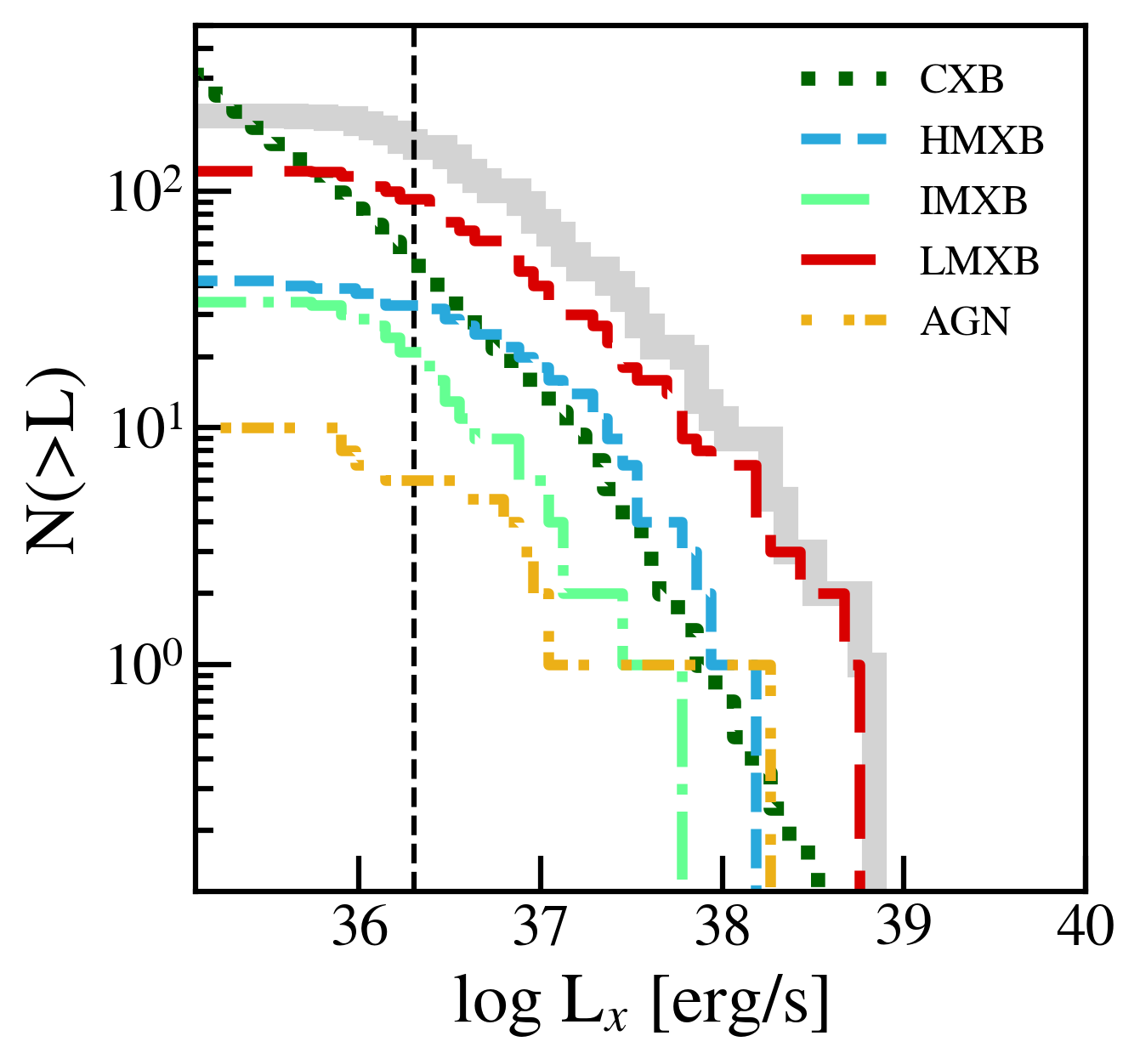}
    \caption{The contribution of X-ray sources to the total cumulative X-ray luminosity function of M81 (grey). \lms\ (red), \ims\ (light green), \hms\ (blue), and AGN (orange) are identified based on the radial profiles and photometry of the optical counterparts as described in \S\ref{sec:sourceclass}. The expected cosmic X-ray background, derived from \cite{kimchamp} using the ChaMP survey fitting parameters, is shown in dark green. The M81 90\% completeness limit of $\ell_X \simgt 36.3$ (from L19) is shown as a vertical dashed line.} 
    \label{fig:cxb}
\end{figure}

\subsection{The Cosmic X-ray Background}\label{sec:cxb}

In Figure~\ref{fig:cxb}, we plot the cumulative XLFs of the M81 X-ray sources classified as  low-, intermediate-, and high-mass XRBs (based on their candidate optical counterparts) within the L19 ellipse, as well as the XLF of the optically identified AGN. These are compared with the expected XLF of Cosmic X-ray Background (CXB) sources, based on the log(N)-log(S) from the ChaMP \cxo\ survey using the $0.5-8$~keV band fit, which has a limiting flux of $6.9 \times 10^{-16} $ ergs cm$^{-2}$ s$^{-1}$ | intermediate between the Chandra Deep Fields and previous surveys \citep{kimchamp}. The ChaMP survey covers an area of 9.6 square degrees, wider than many prior CXB studies, and detected $\sim6500$ sources |the most detected in a single satellite survey. The wider field of view means that cosmic variance will have less an impact on observations than for smaller, deeper fields. At the same time, the flux limit of our X-ray data is shallower than ChaMP's.

Unlike in M83, where the contribution of the CXB is expected to be negligible outside of the $K_s=20$ ellipse, the proximity and larger angular size of M81 imply that the CXB can be expected to contribute significantly to the detected X-ray source population (see also Figure~3 of L19). Specifically 55 CXB sources (that is, AGN) are expected down to the completeness limit within the L19 ellipse (dark green line in Figure~\ref{fig:cxb}), whereas we directly identify the host galaxies of only 10 sources. This implies that optically obscured AGN are a non-negligible source of contamination to the compact X-ray source population within the field of view of M81. We expect those will be mostly associated with LMXB candidates with no detected optical counterparts. 

To test this hypothesis, we conduct a thorough inspection of of 3 \hst\ fields: one covering the outskirts of M81 (outside of the weak disk/L19 ellipse), one covering the weak disk, and one covering the bright disk and a portion of of the bulge. We identify any optical source falling within the 3 fields with a radial profile similar to a galaxy (i.e. non-point source, non-cluster, with extended or unusual shapes, excluding sources that are clearly crowded stars) as a potential galaxy. The aim of this exercise is to quantify the fraction of optically identified galaxies that have an associated X-ray emitting source (AGN), and to assess how that varies as a function of galactocentric distance. The result is a high-end estimate of the number of background galaxies captured within each field.

In the outskirt field, where background galaxies should be virtually unabsorbed and easily identifiable, we identify 545 background galaxies, none of which have an X-ray counterpart. For a field covering 37,913 arcsecs$^{2}$, this translates into a number density of less than 0.02 galaxies per arcsecs$^{2}$, or 1 galaxy for every 70 arcsecs$^{2}$. Across the same field there are 7 compact X-ray sources (4 of which are above the X-ray completeness limit); none are associated with an optical galaxy. 
Owing to increased absorption and crowding, the number of optical galaxies goes down to 349 and 104 for the weak disk and the bright disk/bulge HST fields, respectively; the number of compact X-ray sources in these regions is 7 and 10 (4 and 8 above the completeness limit), respectively. Again, none are identified as AGN. 

This supports our hypothesis that the X-ray emitting AGN that make up the CXB over the field of view of M81 have no detectable optical counterparts, and are thus highly likely to be `hidden' within the objects that we classify as LMXBs. This, combined with our chance superposition assessment analysis (\S\ref{sec:chance}) indicates that our HMXB population suffers from minimal (if any) CXB contamination, whereas the LMXB population is likely contaminated, particularly at the low luminosity end. Specifically, out of the 82 LMXBs that we identified down to the completeness limit (71 of which are field XRBs) within the regions mostly likely to be obscured (i.e. within the outer bounds of the weak disk), a large fraction (of the order of 50) are expected to be potential AGN. 

        \begin{figure*}[t]
            \centering            \includegraphics[width=0.35\linewidth,angle=90,origin=c]{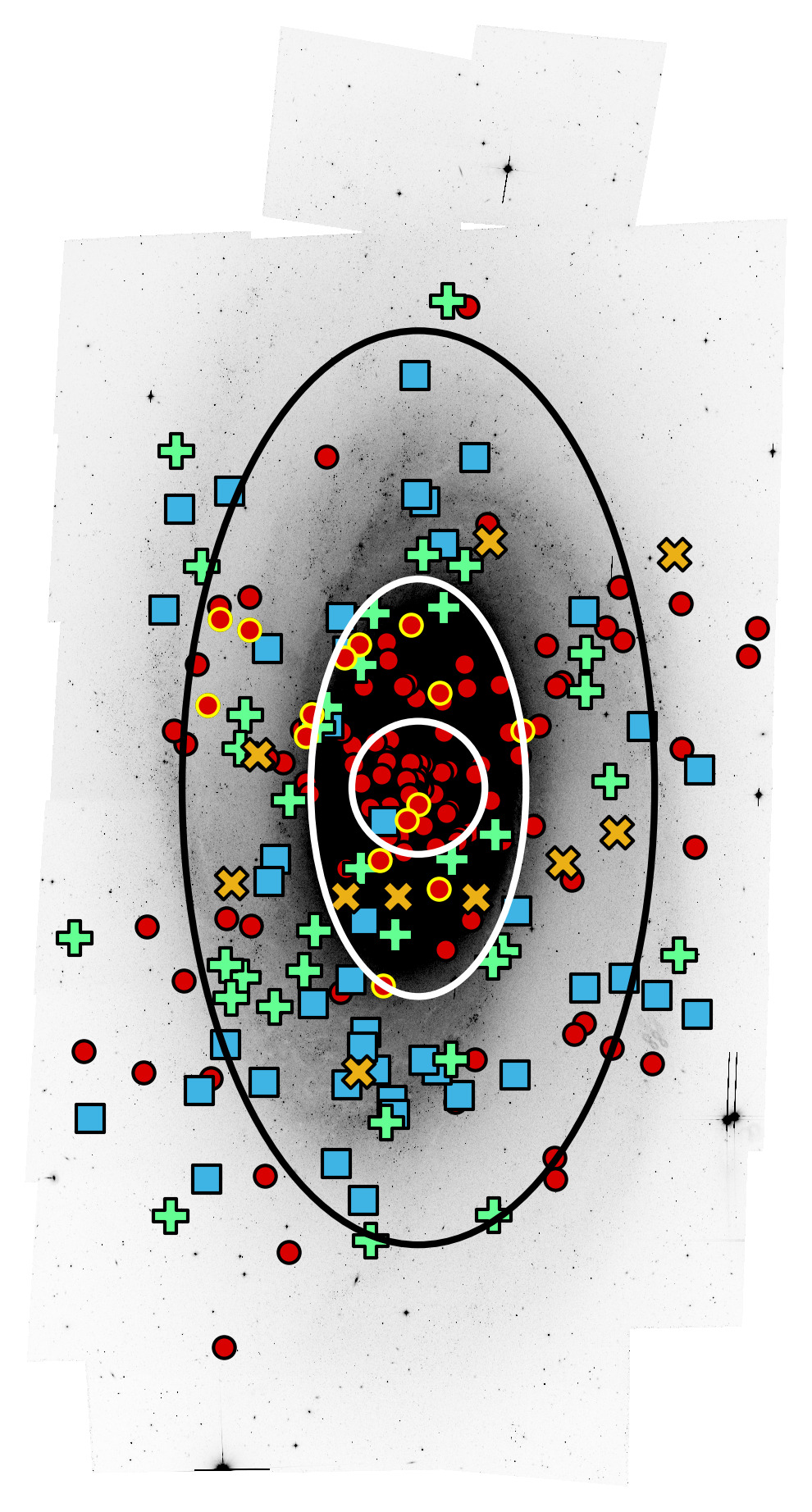}
            \vspace{-2.0cm}
            \caption{Spatial distribution of LMXBs (red points), with globular cluster hosts circled in yellow, IMXBs (green crosses), and HMXBs (blue boxes) on an inverted image of M81. Background galaxies are also shown (orange X's). The 4 regions (bulge, bright disk, weak disk, and outskirts) are marked by black and white ellipses.
            }
            \label{fig:sd}
        \end{figure*}

        \begin{figure*}[t]
        \centering
           \includegraphics[width=.35\linewidth,angle=90,origin=c]{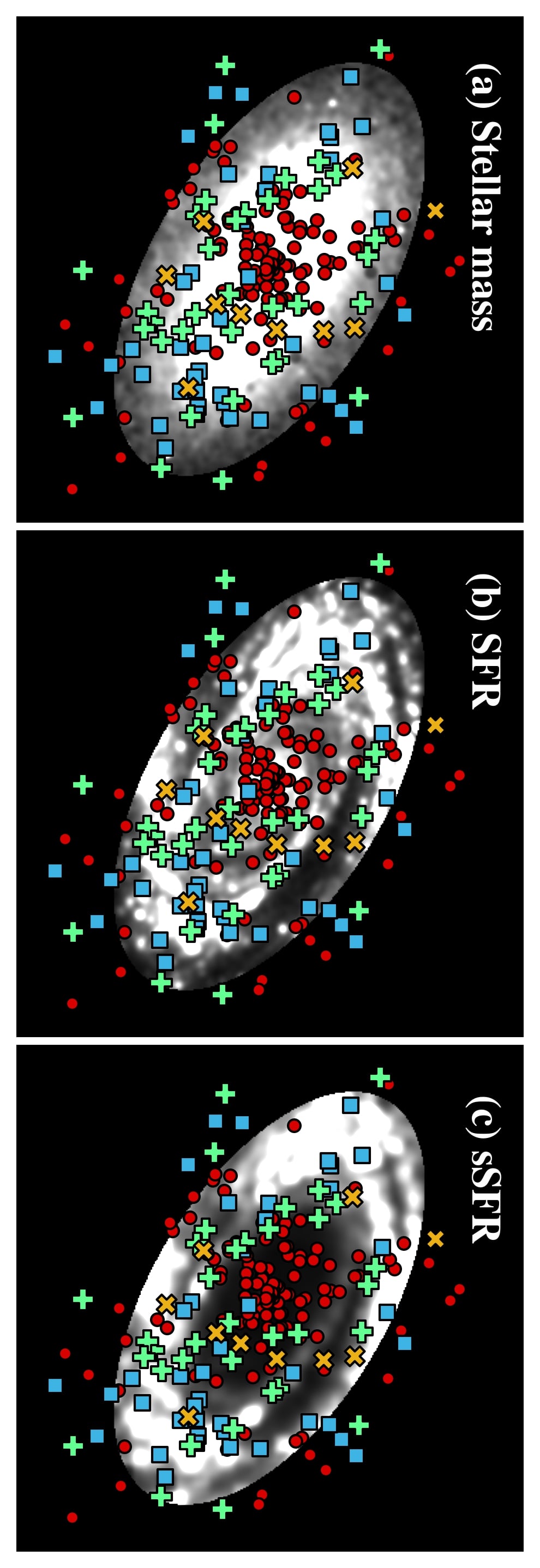}
        \vspace{-6.0cm}
        \caption{Overlays of LMXBs (red points), IMXBs (green crosses), HMXBs (blue squares), and background galaxies (orange X's) onto the (a) stellar mass, (b) star formation rate, and (c) specific star formation rate maps for M81 generated by L19 within the L19 ellipse. All three maps are shown with a linear color scale. \label{fig:lehmers}}
        \end{figure*}

\begin{figure}
    \centering
    \includegraphics[width=\linewidth]{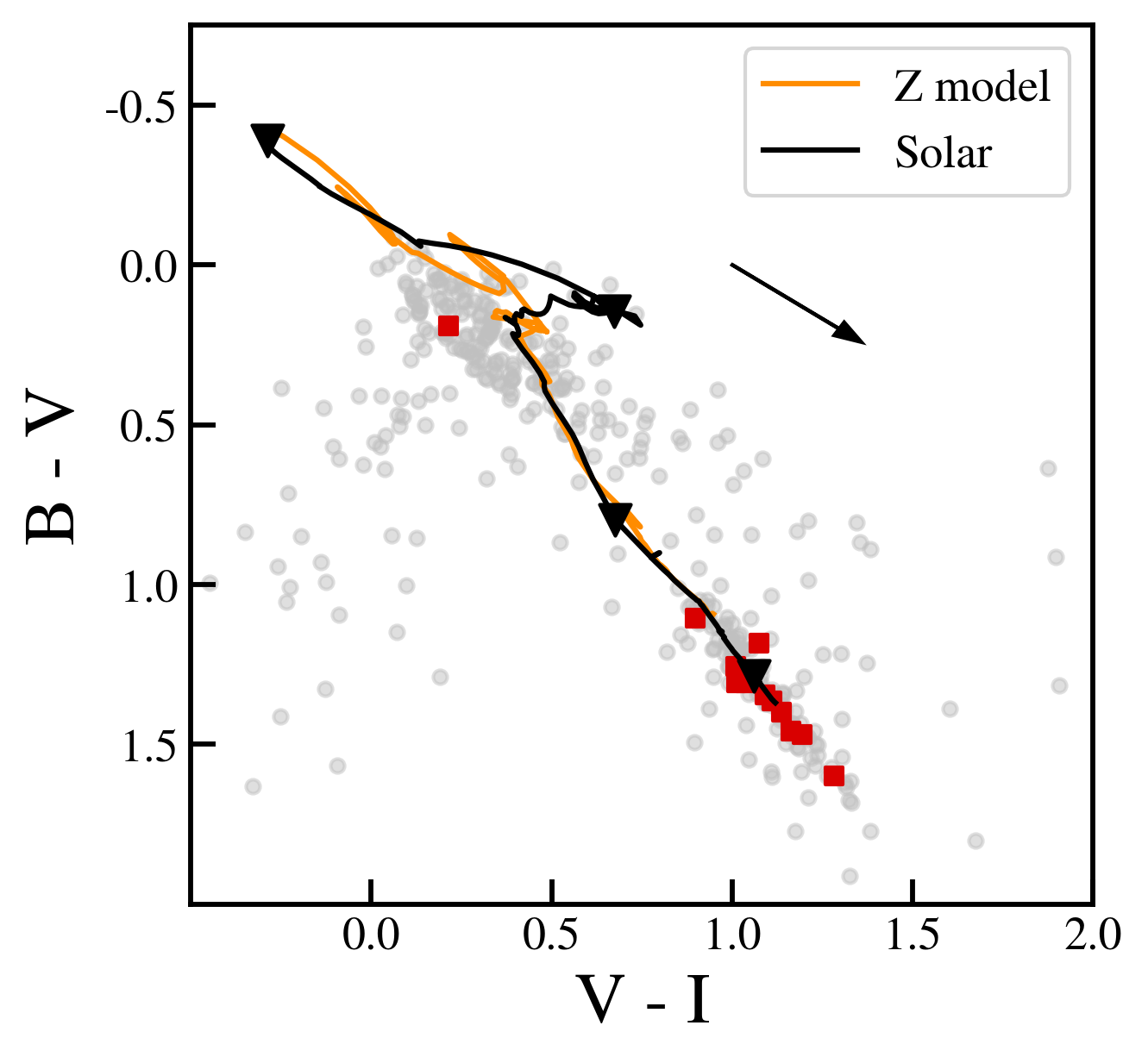}
    \caption{The color-color diagram of clusters from the SC10 catalog, compared to those that we determine are XRB hosts. The black and orange lines represent the \cite{bruzual03} cluster evolution models for solar and 20\% solar metallicities, respectively. Cluster colors are taken from the SC10 measurements. The arrow represents the direction of reddening following the Milky Way extinction law with an example reddening of $E(B-V)=0.2$.}
    \label{fig:sc10_ccd}
\end{figure}

\begin{figure*}
    \centering
    \includegraphics[width=.29\textwidth]{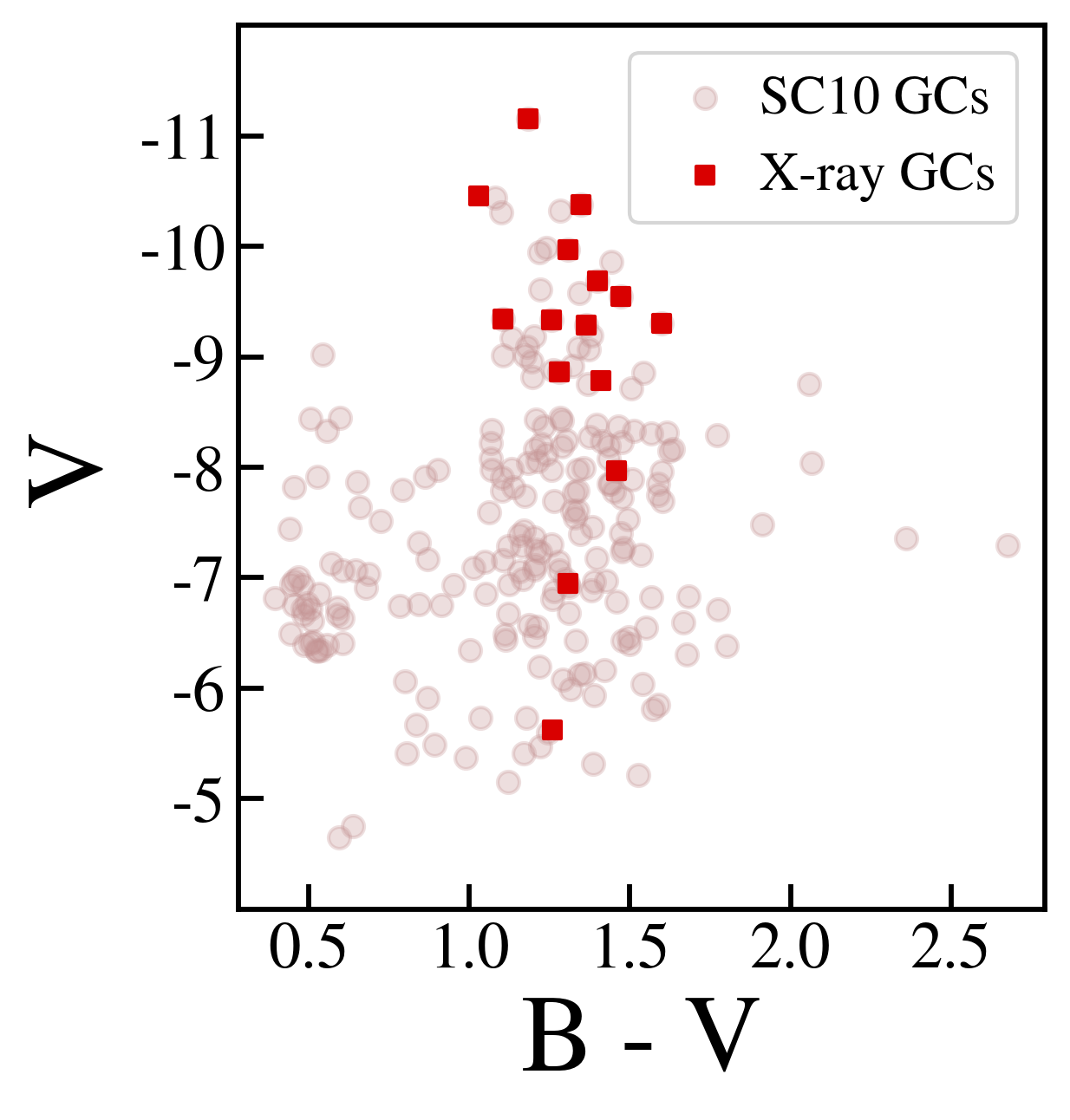}
    \includegraphics[width=.706\textwidth]{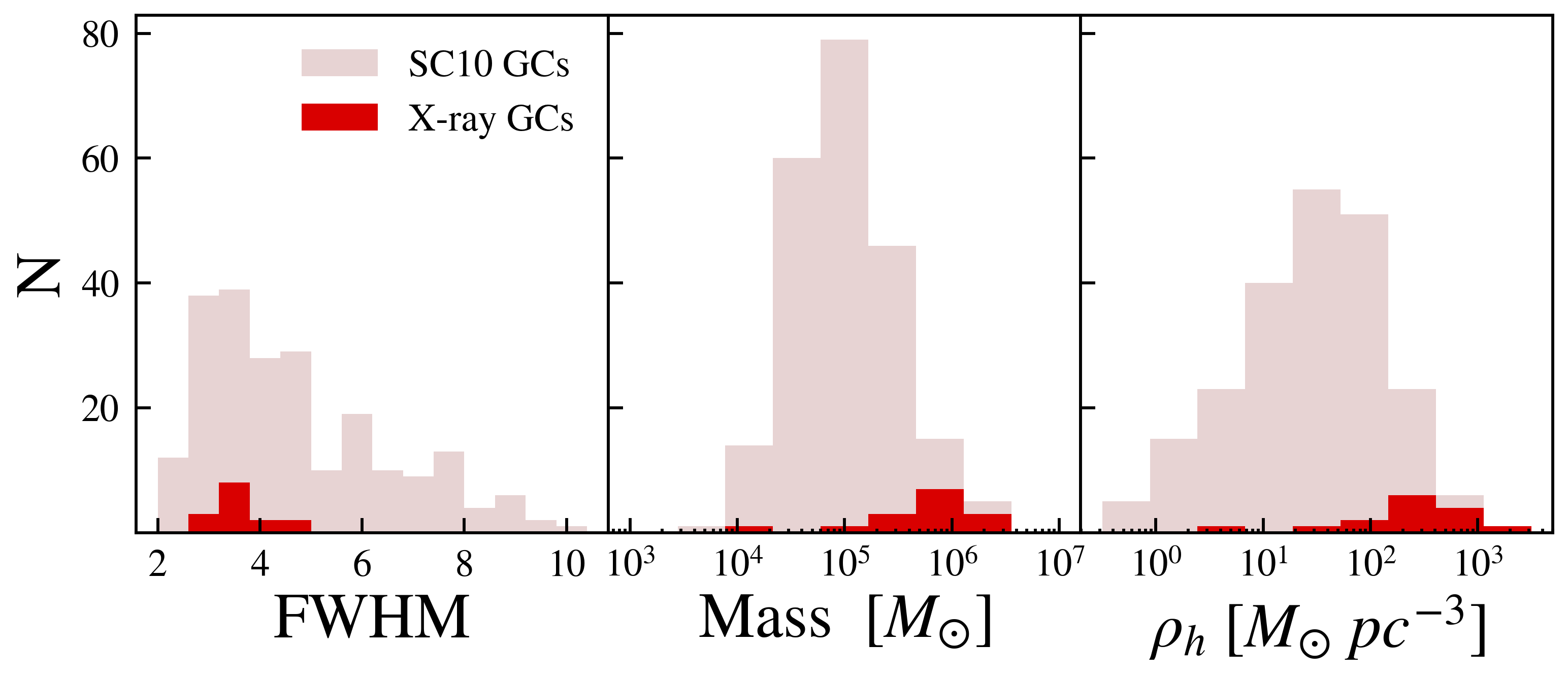}
    \caption{Comparing the properties of X-ray emitting SC10 clusters to all GCs in the SC10 catalog. The left shows the color magnitude diagram, with red squares representing GCs that are XRB hosts. The magnitudes are taken from the SC10 measurements. On the right, the radii and densities of the XRB-hosting GCs are compared to the total GC population.}
    \label{fig:sc10_gcs}
\end{figure*}

\begin{figure*}[t]
    \centering
    \includegraphics[width=0.5\linewidth]{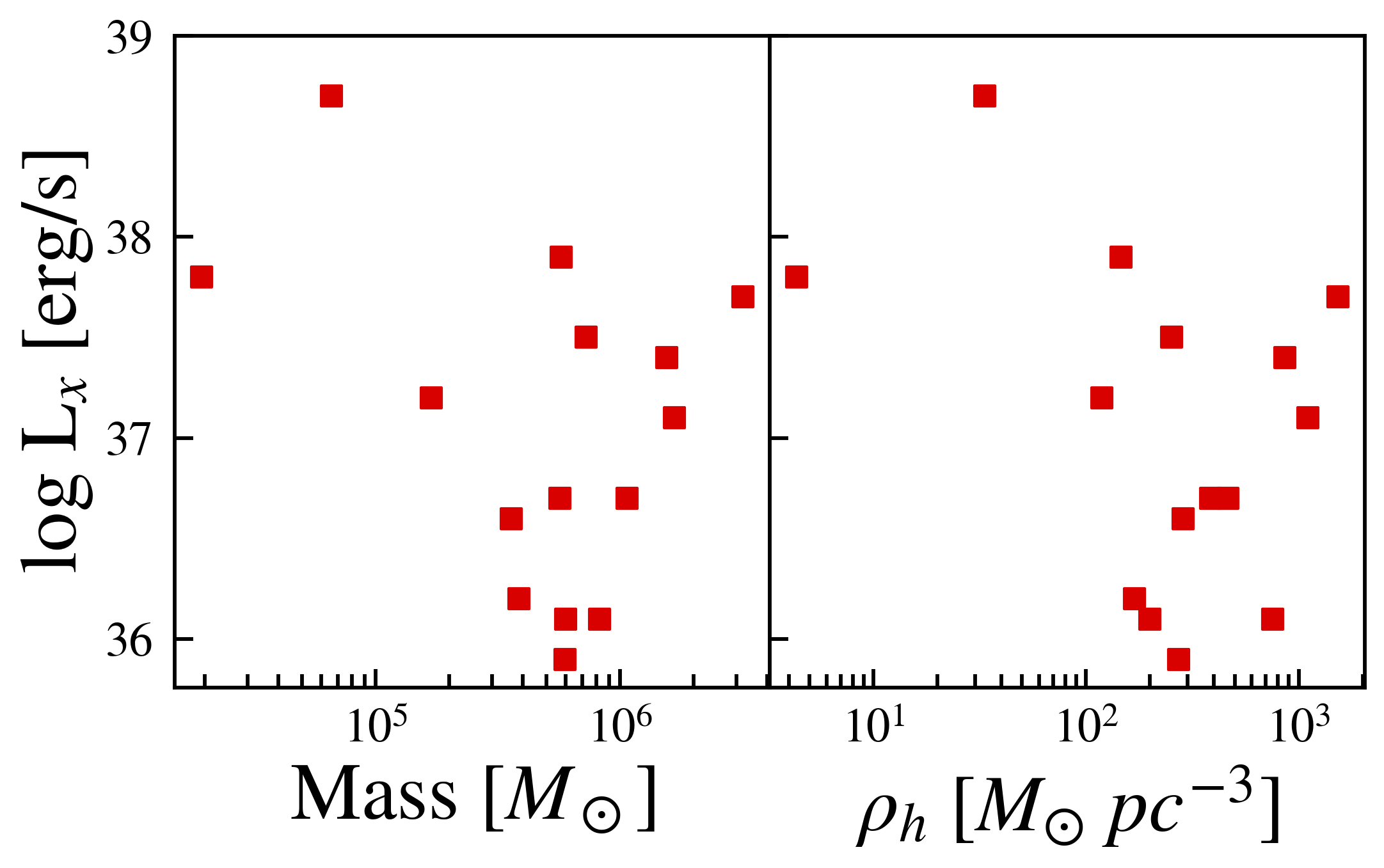}
    \includegraphics[width=0.29\linewidth]{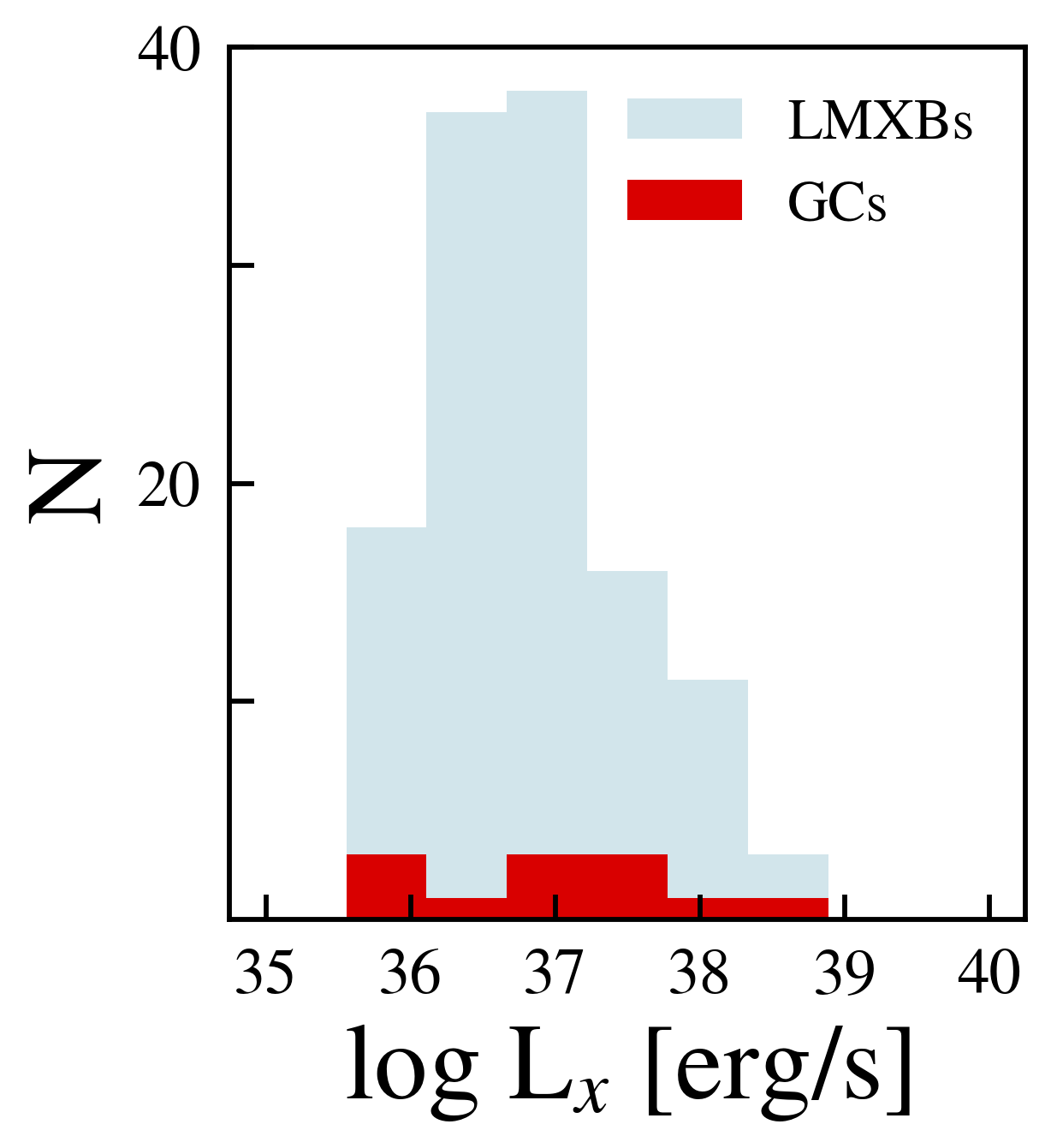}
    \caption{The X-ray properties of the SC10 XRB-hosting clusters. On the left, histograms of the densities and masses of the XRB-hosting GCs from the SC10 catalog are compared to their L19 X-ray luminosities. On the right, a histogram of the X-ray luminosities of the SC10 clusters is compared to our total M81 \lm\ population, not accounting for possible misclassifications (see \S\ref{sec:chance}).}
    \label{fig:sc10_xray}
\end{figure*}

\section{Discussion}

\subsection{LMXB Production in Globular Clusters}\label{sec:gcs}

Extensive research has been done into the population of \lms\ found within GCs in \textit{elliptical galaxies}, in large part due to the abundance of GCs and the ease with which \lms\ can be isolated in galaxies with old stellar and cluster populations. Several early works found that GCs are especially efficient at forming \lms\ thanks to the increased likelihood of dynamical interactions within dense environments. Up to 70\% of all \lms\ within any given early-type galaxy are found within old star clusters, and some particularly luminous GCs appear to harbor more than one \lm\ \citep{angelini01,jordan04, kundu07}. Furthermore, a significant fraction of field \lms\ may have formed originally within GCs, only to be ejected later or deposited within the field when the parent GC dissolved \citep{kundu03, lehmer20}.

There is evidence to suggest metal-rich GCs form \lms\ 2-4 times more efficiently than their metal-poor counterparts \citep{jordan04,kundu07,humphrey08}. One explanation for this may be that, because they have larger radii, metal-rich stars are better able to facilitate both the formation of binaries and the Roche-lobe overflow that characterizes \lms\ \citep{bellazzini95}. Alternatively, this effect may be explained by the potentially longer lifetimes of metal-rich \lms\ due to stronger stellar winds (and therefore faster evolution) of lower-metallicity stars \citep{maccarone04}. On the other hand, metal-rich clusters tend to be more compact, with may lead to more dynamical interactions within the cluster environment \citep{Peacock10}. Despite these hypotheses, a suitable explanation for this apparent dependence is still a subject of much debate.

For all the uncertainty regarding the role of GCs in XRB production in early-type galaxies, even less is known about their role in \textit{late-type galaxies}. Spiral galaxies tend to host GC populations with a range of ages, which makes it difficult to pursue the effect of metallicity on XRB populations due to the degenerate effects of both age and metallicity on cluster colors \citep{kundu03}. In our previous work, we found only 1 \lm\ in a GC in M101 \citep{chandar20} and 4 within M83 \citep{hunt21}. These are much lower fractions of the total \lm\ population than expected in early-type galaxies, since massive spiral galaxies like M83 and M101 are expected to have on the order of $\sim100$ GCs.
Given that M81 is a spiral galaxy with a low SFR (0.25~\msun\textrm{yr}$^{-1}$) and relatively high \sn\ (1.1), it is the ideal target for analyzing the production of XRBs within the GCs of late-type galaxies. Furthermore, the clusters in M81 have been well-studied, allowing us to analyze the properties of XRB-hosting clusters compared to non-XRB clusters and, along the same lines, the differences between field XRBs and cluster XRBs using previously published works. 

As a primary basis for our analysis we use an M81 cluster catalog by \citet[][hereafter SC10]{sc10}. Within this work, a total of 435 compact stellar clusters were identified using \hst\/ACS F435W, F606W, and F814W data in combination with the source detection code SExtractor. They used bounds of $2.4 <$ FWHM $< 10$ ACS pixels (or $2.1 - 9$ pc) for their definition of a compact cluster. 10-pixel aperture magnitudes were also generated by SExtractor, from which $B-V$ and $V-I$ colors are derived. 

Of the 435 clusters in the SC10 catalog, 13 coincide with X-ray sources within our catalog. Of these, 12 are located in GCs, and 1 (located in a young cluster) is identified as a potential SNR by our methods outlined in \S\ref{sec:snr}, which we remove. We also find an additional 3 GC \lms\ not included in the SC10 catalog, which suggests that the SC10 catalog is not complete. To account for these 3 missing GCs in our analysis of the cluster population in M81, we compare the measured properties of the X-ray clusters in SC10, as measured by SExtractor, to the values we measure by our methodology, as described in \S\ref{sec:optcount}. We find that, between the two catalogs, the FWHM measurements for the clusters have a median difference of 0.33 pixels, and the V-band photometry has a median offset of approximately 0.09 mags, while our B-band photometry is fairly consistent with the SC10 measurements. Thus, by applying these offsets to our 3 GCs that are missing in the SC10 catalog, we are able to calibrate our measured values to the SC10 system, which will enable us to include our full GC \lm\ population in our analysis of M81 clusters. 

Figure \ref{fig:sc10_ccd} shows all 435 SC10 clusters (grey points) compared to two models for cluster evolution from \citet{bruzual03}: a solar metallicity model, and a model with metallicity $Z=0.004$, or 20\% solar. As is the case for Figure \ref{fig:cluster}, cluster ages increase down and to the right, and the arrow represents the direction of reddening for M81. The red squares represent X-ray emitting clusters that appear in our source catalog, i.e. XRB hosts. All but one fall towards the lower-right end of the model track, where we expect GCs.  

Since we are interested in understanding the population of LMXBs  within GCs, we isolate the GCs in the SC10 sample by making a color cut corresponding to minimum cluster age of 400 Myrs, based on the \citet{bruzual03} models. 220 of the 435 clusters meet these requirements. This cut successfully selects the 12 clusters we classify as \lm-hosts in our sample, which corresponds to roughly 4.4\% of the total GC population in SC10. The additional 3 GCs not present in the SC10 catalog are also added to the total X-ray GCs population for the remainder of our analysis. We calculate the masses of the GCs by adopting a mass-luminosity ratio of 1.5~\msun/\lsun\ as presented in \citet{chandar07}. The mass of each cluster is estimated by converting their V-band magnitudes using an absolute Vega magnitude of 4.66 for the Sun, as viewed in the ACS F606W filter \citep{willmer18}. From these masses, we also calculate cluster densities following \cite{chandar07}, converting FHWMs to effective radii using a pixel scale of 0.\arcsec05~$\mathrm{pixel}^{-1}$ and a conversion factor of 1.48 \citep{anders06,mulia19}.

In Figure \ref{fig:sc10_gcs}, we compare the properties of GCs that host XRBs (red) with the properties of those that do not (pink) | that is, the 15 X-ray emitting GCs from our sample compared to the 208 GCs from the SC10 catalog that do not have X-ray counterparts. We find that, within the spiral galaxy M81, GCs that host XRBs tend to have higher V-band magnitudes than non-XRB GCs, as demonstrated in the leftmost panel. Furthermore, when observing the sizes, masses, and densities of the clusters (right panels), we find that \textit{GCs that host XRBs have smaller radii, are more massive, and hence denser than GCs that do not}. 
Higher masses and densities may be more conducive for the formation of XRBs, since dense environments allow for more close interactions between stars, increasing the chances of capture. 
Similarly, higher masses correlate with more stars and, therefore, more opportunities for dynamical interactions \citep{Peacock10}.
\par 
With only BVI photometry, it is hard to make definitive statements about the metallicities of GCs in spiral galaxies, since small amounts of reddening would significantly change the estimated metallicity.  We do find however, that the majority of GCs which host LMXBs in M81 are more centrally concentrated than the full SC10 GC catalog.  This is potentially indirect evidence that LMXBs may prefer to form in more metal-rich GCs in spiral galaxies as well, since metal-rich GCs tend to be associated with bulges (rather than halos), and more concentrated towards the centers of their host galaxies as a result.

When we analyze the properties of the 15 X-ray emitting GCs (Figure \ref{fig:sc10_xray}, left), we find no significant correlation between X-ray luminosity and the mass or density of the host cluster. That is, more massive XRB-hosting GCs are \textit{not} more luminous in the X-rays, in line with what was found by \citet{sivakoff07} for GCs in Virgo cluster elliptical galaxies.
It is also interesting to compare the population of \lms\ within GCs to the total number of XRBs and to field \lms\ specifically, although the latter is quite uncertain in M81. Of the 199 XRBs we identified within M81 (excluding SNR candidates), 15 are found in GCs. 159 XRBs, 11 in GCs, are above the completeness limit.  Taken at face value, we classified 122 XRBs as low-mass (100 above the completeness limit).  However, as discussed in \S\ref{sec:cxb}, this population likely has significant contamination from CXB sources, and may be missing other sources classified as \ims\ (due to, e.g., intense X-ray irradiation).
The X-ray luminosities of XRB-hosting GCs are consistent with field XRBs, independent of whether they are compared to \textit{only} low-mass field XRBs (blue sources in the right panel of Figure \ref{fig:sc10_xray}) or to the \textit{total} field XRB population. This may bolster the idea that GCs are ``seeding" the field population of \lms\ \citep{lehmer20}. 

Overall, our observations of XRB-hosting GCs in M81 is consistent with observations of those in early-type galaxies. However, as we noted, a more complete sample of GCs in M81 is needed before any strong conclusions may be drawn as to what fraction of GCs may host XRBs in star-forming galaxies.

\begin{figure*}
    \centering
        \centering
        \includegraphics[width=0.45\linewidth]{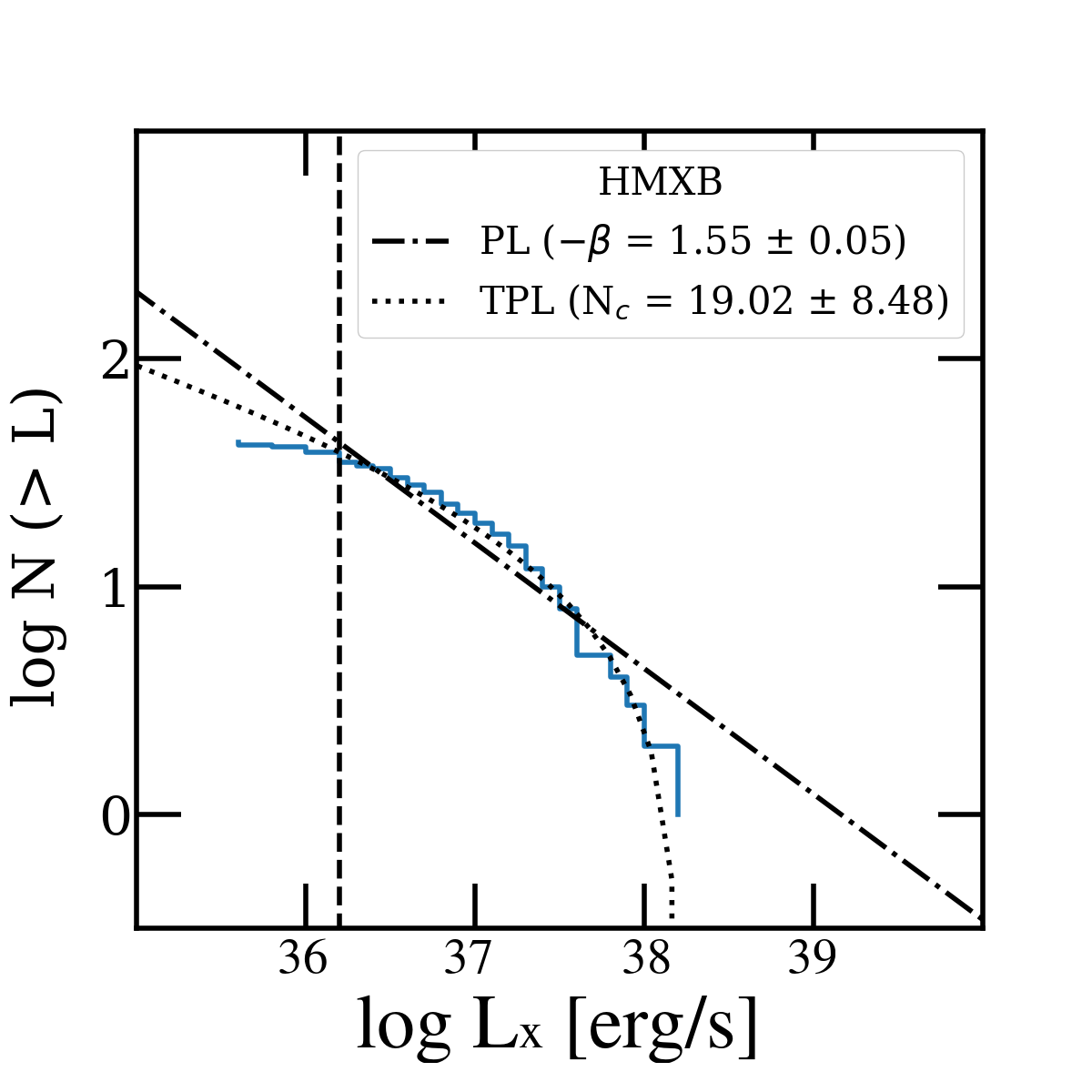} 
        \includegraphics[width=0.45\linewidth]{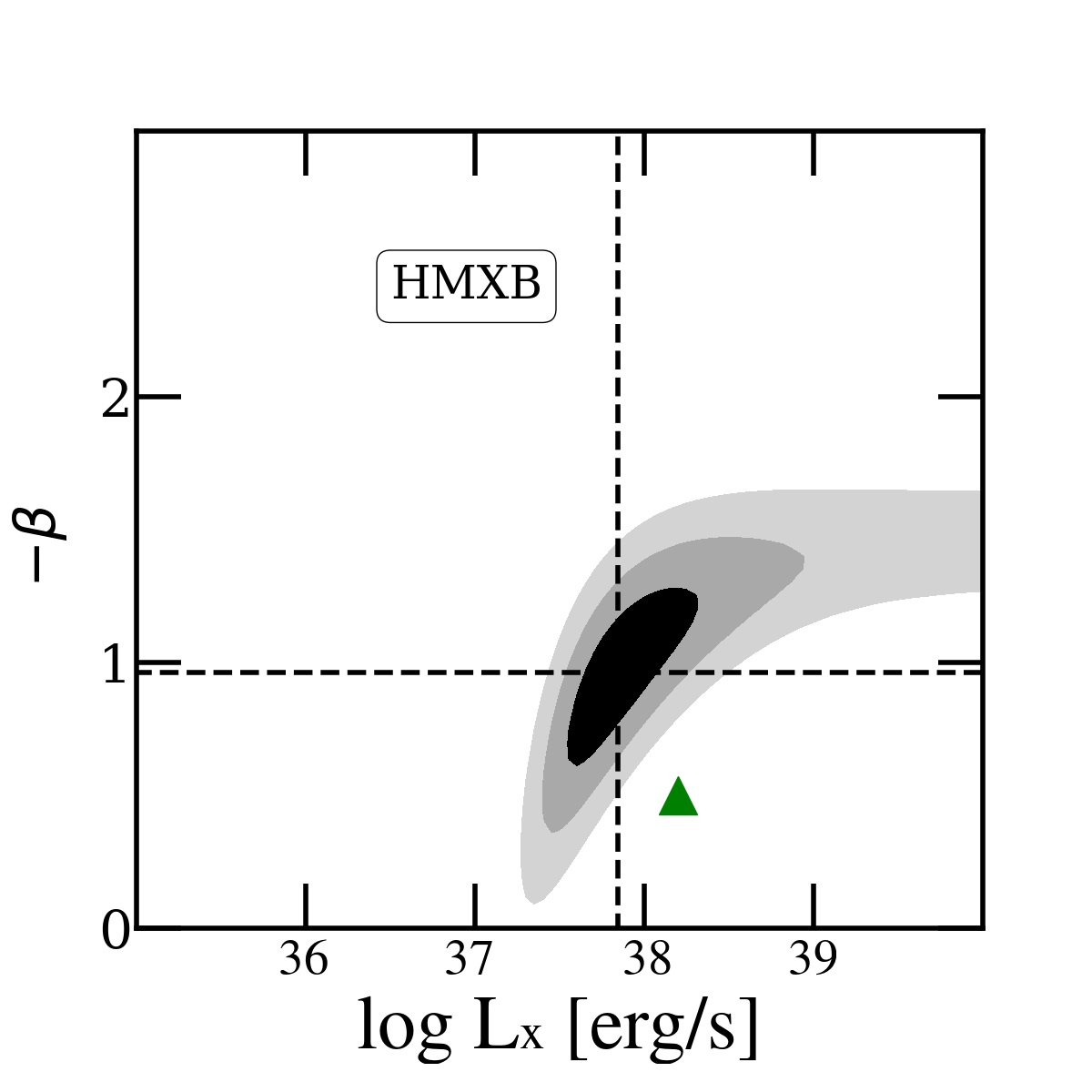} 
    \caption{Left: Fits to the \hm\ XLF of M81 with a cumulative, un-binned power-law function, with and without a truncation at high luminosities (dotted-dashed and dotted lines, respectively). The functions are fitted down to the completeness limit, $\ell_X = 36.3$ (dashed line). Right: Maximum likelihood fit to the cumulative \hm\ XLF using a Schechter function. The green triangle marks the X-ray luminosity of the brightest \hm. Contours represent the 1-, 2-, and 3$\sigma$ confidence level. The dashed lines represent the `best fit' at $-\beta=0.96$ and a knee log-luminosity of 37.84. }\label{fig:hm_fits}
\end{figure*}

\subsection{HMXB XLF in M81}\label{sec:hmxlf}

In this section we will focus on the shape and normalization of the XLF of the HMXB population, which our experiments in \S\ref{sec:chance} show have negligible chance superposition/background contamination. We are especially interested in how our method for classifying \hms\ compares to state-of-the-art X-ray based results (e.g., L19) both in terms of shape as well as normalization. 
Following the analysis conducted in \citet{hunt21}, we examine the \hm\ XLF via two methods: first, we fit the un-binned,  cumulative XLF to a power-law (PL), with and without a high-luminosity truncation, using the IDL script \texttt{MSPECFIT} \citep{mspecfit}. Second, we perform a Maximum Likelihood (ML) fit of the differential XLF to a Schechter function. The former enables us to potentially detect a break (regardless of its functional shape) at high luminosities, which may be expected to be present near the Eddington luminosity for a stellar-mass compact object. The latter is a robust test for the presence of an \textit{exponential} break \citep{mok19}. 

Figure~\ref{fig:hm_fits} shows the results of these fits. The left panel shows the PL fit to the cumulative \hm\ XLF down to the 90\% completeness limit (vertical dashed line). The index of the PL fit is given by $-\beta$, while a value $N_{c} > 1$ for the truncated PL indicates a statistically significant high-energy downturn. The right panel of Figure~\ref{fig:hm_fits} shows the results of the ML fit, where the best-fit index ($-\beta$) is plotted against the best fit Schechter knee luminosity, with the contours representing the 1-, 2- and 3$\sigma$ confidence levels, and the green triangle marking the luminosity of the brightest \hm. The upper limit to the knee luminosity is set to be well in excess of the brightest \hm\ to ensure convergence of the fit.

The results of both fitting methods suggest a marginal cutoff or downturn at a luminosity just below $\ell_X = 38$, as indicated by $N_{c} = 19.02 \pm 8.48$ (left) and a closed $2\sigma$ but an open $3\sigma$ ML contour (right) with a fit of $-\beta = 0.96$ and $\ell_{\rm{knee}} = 37.84$. This is similar to our results for M83, in which only the \hm\ population shows marginal evidence of a downturn (at a $\sim2.5\sigma$ level), albeit at a higher luminosity ($\ell_X \simgt 38$). In contrast, we found that none of the XRB populations in M101 (high-, intermediate-, or low-mass) required a truncation or Schechter-like cutoff \citep{chandar20}. We note, however, that we did not excise the SNR population using the X-ray luminosity and color cuts as for M81 and M83, so it is possible that the inclusion of a large population of low-luminosity SNRs in the XLFs of M101 washed out any PL deviation at high-luminosity. 

We compare our PL fit for \hms\ to that found by L19. Using sub-galactic SFR and \mstar\ maps (the same shown Figure~\ref{fig:lehmers}), they construct both a global \hm\ XLF model that scales with enclosed SFR and a `standard' XLF fit for each individual galaxy in their sample (38 galaxies in total). These fits employ a forward-fitting model that fits both the XRB and CXB contributions simultaneously. They find that the individual galaxy \hm\ XLFs are best fit with a single PL with a fixed high-energy cutoffs of $\ell_{c} = 40.3$. For M81, L19 reports a best fit PL index of $-1.43 \pm 0.06$, which is consistent within the errors with our fitted value of $-\beta = 1.55 \pm 0.05$.

The L19 global \hm\ XLF model gives a slightly steeper index of $-1.66 \pm 0.02$. We can use the global model to estimate how many \hms\ are expected within the L19 ellipse, given the enclosed SFR provided by the maps. L19 gives a global \hm\ normalization of $2.06^{+0.16}_{-0.15}~(M_{\star} {\rm yr}^{-1})^{-1}$, with a galaxy-specific scaling factor of $\omega = 0.90$ for M81. We estimate a total SFR of 0.25 \mstar $\rm{yr^{-1}}$ enclosed within the L19 ellipse (this excludes 0.01 \mstar $\rm{yr^{-1}}$ within 12 arcsec of the bulge, which hosts a bright AGN). Based on these numbers and the functions given by Equation 4 in L19, 10 \hms\ with $\ell_X \ge 36.3$ are expected within the bounds of the L19 ellipse. This is to be compared with 26 \hms\ classified by our optical methods. As discussed in \S\ref{sec:misclass}, these are highly likely to be genuine counterparts as the chance superposition with a high-mass star is negligible across the whole disk. 

\section{Summary and Conclusions}
 
We make use of multi-wavelength \hst\ imaging data to characterize the compact X-ray source population of M81 (using the X-ray source catalog by \citealt{lehmer19}), a nearby spiral galaxy with moderately low SFR. We directly characterize each X-ray source as a foreground star, a background galaxy/AGN, a SNR (based on the X-ray luminosity/hardness criterion developed by \citealt{hunt21}) or an XRB. In the latter case, we use color-magnitude diagrams to tentatively identify the most likely donor as a low-, intermediate- or high-mass star, or colors to estimate age in cases where the XRB is associated with a stellar cluster.
In summary, we find the following: 

\begin{itemize}
    \item 199 out of 240 X-ray sources that are found over the HST footprint of M81 are classified as XRBs (Figures~\ref{fig:mosaic}~and~\ref{fig:sd}). Out of these, 159 are above the X-ray completeness limit of $\ell_X~>~36.3$. Based on color-magnitude diagrams of the most likely counterpart(s), we identify 100 \lms\ (no visible counterpart), 25 \ims\ (donor mass between 3-8 \msun), and 34 \hms\ (donor mass above 8 \msun). 
    
    \item After a thorough  misclassification assessment using artificial sources to quantify chance superpositions, we conclude that the \hms\ population is robust, with a very low probability that a X-ray source will have a chance superposition with a massive star. In contrast, X-ray sources with no detectable counterparts likely harbor a significant population of background AGN with no visible optical host galaxy, whereas \ims\ suffer from a non-negligible likelihood of chance superposition (additionally, intense X-ray irradiation may contribute to enhance the brightness and alter the colors of genuinely low-mass donors). 
    \item 15 XRBs are found in old GCs, whereas 2 are associated with young star clusters ($<400$ My old). 
    GCs that host XRBs tend to be more massive and more compact than GCs that do not, similar to earlier findings in elliptical galaxies. These have X-ray luminosities consistent with the field population, and there is no obvious correlation between the X-ray luminosities of XRBs in GCs and the masses or densities of their host clusters.
    \item The XLF of \hms\ in M81 shows a marginal high-luminosity break at $\ell_X = 37.84$, similar to that found for M83 \citep{hunt21}. We note that the emergence of such a cut-off around the Eddington limit for a stellar-sized accretor may depend on whether the contamination from SNR at the low-luminosity end of the XLF is properly accounted for. While this effect may be negligible in a low star forming galaxy, it is likely to affect the XLFs of late types with high SFRs (such as M83).  

\end{itemize}

\begin{acknowledgements}
QH is partially funded by a Rackham Merit Fellowship, awarded by the University of Michigan Rackham Graduate School. QH thanks the LSSTC Data Science Fellowship Program, which is funded by LSSTC, NSF Cybertraining Grant \#1829740, the Brinson Foundation, and the Moore Foundation; her participation in the program has benefited this work. We are also grateful to Xuheng Ding, Tommaso Treu, and Tom Maccarone for useful comments and suggestions.
\end{acknowledgements}

\software{AstroPy (The Astropy Collaboration 2013, 2018), DrizzlePac (Avila et al. 2014), IRAF (Tody 1986, Tody 1993), Matplotlib (Hunter 2007), MSPECFIT (Rosolowsky 2005), SAOImage DS9 (Joye \& Mandel 2003)}

\newpage
\bibliography{Hunt_M81_source}

\appendix{\input{Sources_table}
\input{mosaics}}

\end{document}

%% file: Sources_table.tex
\startlongtable
\begin{deluxetable*}{cccccccccc}
\tabletypesize{\footnotesize}
\tablewidth{0pt}
\tablecaption{Properties and Classifications of M81 X-ray Sources }\label{tab:allsources}
\tablehead{     
\colhead{ID} & \colhead{CSC ID} &  \colhead{R.A.} &  \colhead{Dec} & \colhead{\lx} &          \colhead{V} &    \colhead{B-V} &   \colhead{V-I}& \colhead{Class} & \colhead{CF}}
\startdata
    M81X003 & | & 148.579 & 69.07 & 36.5 & -2.393 & | & 1.199 & LM & 1.0 \\
    M81X006 & 2CXO J095422.6+690346 & 148.595 & 69.063 & 37.1 & -0.552 & | & 2.552 & LM & 1.0 \\ 
    M81X007 & | & 148.618 & 69.099 & 36.6 & -4.616 & 0.382 & 0.694 & Gal & 1.0 \\ 
    M81X008 & 2CXO J095431.1+690506 & 148.629 & 69.085 & 36.6 & -1.636 & | & 1.497 & LM & 1.0 \\ 
    M81X009 & 2CXO J095434.5+690452 & 148.644 & 69.081 & 36.7 & -5.612 & 1.194 & 1.045 & (SNR) & 1.0 \\ 
    M81X010 & 2CXO J095436.0+690856 & 148.651 & 69.149 & 37.0 & -2.934 & | & 1.389 & (SNR) & 1.0 \\ 
    M81X011 & | & 148.654 & 69.078 & 36.4 & -8.706 & 1.875 & 0.520 & Star & 1.0 \\ 
    M81X013 & | & 148.669 & 69.038 & 36.2 & -4.473 & 2.054 & 0.976 & HM & 1.0 \\ 
    M81X014 & 2CXO J095440.8+690548 & 148.67 & 69.097 & 36.4 & -3.081 & 1.997 & 1.263 & LM & 2.0 \\ 
    M81X015 & 2CXO J095442.0+690244 & 148.675 & 69.046 & 37.0 & -1.591 & | & 1.552 & LM & 1.0 \\ 
    M81X016 & 2CXO J095444.3+690454 & 148.685 & 69.082 & 37.0 & -2.633 & | & 1.636 & LM & 1.0 \\ 
    M81X018 & 2CXO J095446.4+690513 & 148.694 & 69.087 & 37.0 & -2.5 & | & 1.650 & LM & 1.0 \\ 
    M81X019 & 2CXO J095446.7+691123 & 148.695 & 69.19 & 37.1 & -2.169 & | & 1.135 & LM & 1.0 \\ 
    M81X020 & 2CXO J095447.2+690101 & 148.697 & 69.017 & 36.8 & -2.712 & 1.421 & 2.017 & LM & 1.0 \\
    M81X021 & 2CXO J095447.4+690322 & 148.698 & 69.056 & 36.2 & -8.6 & -0.009 & 0.658 & \textit{HM} & 1.0 \\ 
    M81X022 & 2CXO J095449.2+690538 & 148.705 & 69.094 & 36.5 & -5.155 & 0.362 & 0.365 & HM & 1.0 \\
    M81X023 & | & 148.708 & 69.194 & 37.2 & -3.142 & | & 1.816 & IM & 1.0 \\ 
    M81X024 & 2CXO J095452.0+690455 & 148.717 & 69.082 & 36.4 & -3.703 & | & 1.954 & IM & 1.0 \\ 
    M81X025 & 2CXO J095455.0+690419 & 148.73 & 69.072 & 36.2 & -3.349 & | & 1.826 & IM & 1.0 \\ 
    M81X026 & 2CXO J095455.8+690517 & 148.733 & 69.088 & 36.3 & -4.379 & 0.423 & 0.256 & (SNR) & 1.0 \\ 
    M81X027 & | & 148.738 & 69.149 & 36.5 & -5.269 & -0.067 & -0.168 & HM & 1.0 \\ 
    M81X028 & 2CXO J095457.6+690241 & 148.74 & 69.045 & 37.5 & -3.578 & | & 0.281 & IM & 1.0 \\ 
    M81X029 & 2CXO J095458.4+685922 & 148.744 & 68.99 & 36.2 & -2.986 & 0.158 & 0.169 & IM & 1.0 \\ 
    M81X030 & 2CXO J095458.8+690521 & 148.745 & 69.089 & 36.1 & | & | & | & LM & 1.0 \\ 
    M81X031 & 2CXO J095458.8+690438 & 148.745 & 69.077 & 36.5 & | & | & | & LM & 1.0 \\ 
    M81X032 & 2CXO J095459.8+685817 & 148.75 & 68.972 & 36.2 & -5.095 & 0.268 & 0.101 & HM & 1.0 \\ 
    M81X033 & 2CXO J095500.0+690745 & 148.75 & 69.129 & 38.3 & -2.514 & 2.16 & | & LM & 1.0 \\ 
    M81X034 & 2CXO J095500.1+690437 & 148.751 & 69.077 & 36.5 & | & | & | & LM & 1.0 \\ 
    M81X035 & 2CXO J095500.3+690148 & 148.752 & 69.03 & 37.1 & -4.37 & 0.841 & 1.937 & Gal & 1.0 \\ 
    M81X036 & 2CXO J095501.0+690726 & 148.754 & 69.124 & 38.3 & -3.293 & | & 2.512 & Gal & 1.0 \\ 
    M81X037 & 2CXO J095501.0+685622 & 148.754 & 68.939 & 37.5 & | & | & | & Star & 1.0 \\ 
    M81X038 & 2CXO J095501.7+691040 & 148.757 & 69.178 & 37.3 & -6.276 & 0.685 & 0.157 & HM & 1.0 \\
    M81X039 & 2CXO J095502.6+685621 & 148.761 & 68.939 & 37.2 & | & | & | & Star & 1.0 \\ 
    M81X040 & 2CXO J095505.6+685852 & 148.774 & 68.981 & 36.8 & -4.358 & 1.842 & 0.942 & HM & 1.0 \\
    M81X041 & 2CXO J095506.3+690405 & 148.777 & 69.068 & 36.8 & | & | & | & LM & 2.0 \\ 
    M81X042 & 2CXO J095507.2+690314 & 148.78 & 69.054 & 35.9 & -4.06 & 0.939 & 1.029 & SNR & 1.0 \\ 
    M81X043 & 2CXO J095507.5+690713 & 148.781 & 69.12 & 35.9 & -7.397 & 0.206 & 0.266 & \textit{IM} & 1.0 \\ 
    M81X044 & 2CXO J095508.8+685722 & 148.787 & 68.956 & 36.9 & | & | & | & Star & 1.0 \\ 
    M81X045 & | & 148.79 & 69.071 & 35.8 & | & | & | & SNR & 1.0 \\ 
    M81X046 & 2CXO J095509.6+690743 & 148.79 & 69.129 & 37.4 & -4.855 & 1.267 & 0.819 & HM & 1.0 \\ 
    M81X047 & 2CXO J095509.7+690407 & 148.791 & 69.069 & 37.1 & -10.368 & 0.94 & 0.526 & \textit{LM} & 1.0 \\ 
    M81X048 & 2CXO J095509.6+690832 & 148.791 & 69.142 & 37.2 & -5.145 & 1.008 & 0.920 & HM & 2.0 \\
    M81X049 & 2CXO J095510.2+690502 & 148.793 & 69.084 & 38.5 & -3.411 & | & 1.713 & LM & 2.0 \\ 
    M81X050 & | & 148.793 & 68.99 & 35.6 & -4.884 & 0.058 & 0.011 & HM & 1.0 \\ 
    M81X051 & 2CXO J095510.6+690843 & 148.795 & 69.145 & 37.3 & -5.372 & 0.515 & 0.411 & HM & 1.0 \\
    M81X052 & 2CXO J095511.8+685748 & 148.799 & 68.963 & 36.9 & -1.538 & | & 2.194 & LM & 1.0 \\ 
    M81X053 & 2CXO J095512.2+690344 & 148.801 & 69.062 & 36.2 & | & | & | & LM & 2.0 \\ 
    M81X054 & 2CXO J095512.4+690411 & 148.802 & 69.07 & 35.9 & -2.529 & | & 2.212 & LM & 1.0 \\ 
    M81X055 & 2CXO J095512.4+690121 & 148.802 & 69.023 & 36.7 & | & | & | & LM & 1.0 \\ 
    M81X056 & 2CXO J095512.6+690141 & 148.803 & 69.028 & 36.0 & -6.722 & 0.408 & 0.252 & Gal & 1.0 \\ 
    M81X057 & 2CXO J095514.1+690740 & 148.809 & 69.128 & 36.3 & -4.388 & 0.216 & 0.282 & IM & 1.0 \\
    M81X058 & 2CXO J095514.5+690641 & 148.811 & 69.111 & 36.4 & -3.462 & 0.625 & 0.799 & IM & 1.0 \\
    M81X059 & 2CXO J095515.2+690230 & 148.813 & 69.042 & 36.6 & -2.752 & | & 1.415 & LM & 1.0 \\ 
    M81X060 & 2CXO J095515.2+690537 & 148.814 & 69.094 & 36.5 & | & | & | & LM & 2.0 \\ 
    M81X061 & 2CXO J095516.6+690512 & 148.819 & 69.087 & 36.4 & | & | & | & LM & 2.0 \\ 
    M81X062 & 2CXO J095518.0+685820 & 148.825 & 68.972 & 36.0 & -2.277 & | & 1.801 & LM & 1.0 \\ 
    M81X063 & 2CXO J095518.2+685930 & 148.826 & 68.992 & 36.4 & -5.193 & 0.615 & 0.860 & HM & 1.0 \\
    M81X064 & 2CXO J095519.6+690732 & 148.832 & 69.126 & 36.3 & -4.12 & 0.65 & 0.799 & SNR & 1.0 \\ 
    M81X065 & 2CXO J095519.9+690351 & 148.833 & 69.064 & 36.5 & | & | & | & LM & 2.0 \\ 
    M81X066 & 2CXO J095521.0+690313 & 148.838 & 69.054 & 36.1 & | & | & | & LM & 2.0 \\ 
    M81X067 & 2CXO J095521.1+685855 & 148.838 & 68.982 & 36.4 & -2.802 & | & 1.394 & LM & 1.0 \\ 
    M81X068 & 2CXO J095521.4+690831 & 148.84 & 69.142 & 36.6 & -5.571 & 1.543 & 0.924 & SNR & 1.0 \\
    M81X069 & 2CXO J095521.7+690345 & 148.841 & 69.062 & 36.6 & | & | & | & LM & 2.0 \\ 
    M81X070 & 2CXO J095521.8+690522 & 148.841 & 69.09 & 37.4 & | & | & | & LM & 2.0 \\ 
    M81X071 & 2CXO J095521.8+690637 & 148.841 & 69.11 & 37.7 & -10.613 & 0.92 & 0.864 & \textit{LM} & 1.0 \\ 
    M81X072 & 2CXO J095521.9+690228 & 148.842 & 69.041 & 36.0 & -3.296 & | & 1.397 & LM & 2.0 \\ 
    M81X073 & 2CXO J095522.0+690518 & 148.842 & 69.089 & 36.7 & -9.921 & 1.234 & 0.775 & \textit{LM} & 1.0 \\ 
    M81X074 & 2CXO J095522.1+690510 & 148.842 & 69.086 & 38.2 & | & | & | & LM & 2.0 \\ 
    M81X075 & 2CXO J095522.7+690237 & 148.845 & 69.044 & 35.8 & -3.98 & | & 1.438 & IM & 2.0 \\ 
    M81X076 & 2CXO J095523.7+685849 & 148.849 & 68.98 & 37.2 & -2.347 & | & 2.225 & LM & 1.0 \\ 
    M81X077 & 2CXO J095524.2+690957 & 148.851 & 69.166 & 38.8 & -2.534 & | & 1.242 & LM & 1.0 \\ 
    M81X078 & 2CXO J095524.2+690439 & 148.851 & 69.078 & 36.7 & | & | & | & LM & 2.0 \\ 
    M81X079 & 2CXO J095524.7+690113 & 148.853 & 69.02 & 38.2 & -7.485 & 1.003 & 0.601 & HM & 1.0 \\ 
    M81X080 & 2CXO J095525.6+690458 & 148.857 & 69.083 & 36.1 & | & | & | & (SNR) & 2.0 \\ 
    M81X081 & | & 148.86 & 69.067 & 36.6 & | & | & | & LM & 2.0 \\ 
    M81X082 & 2CXO J095526.6+690523 & 148.861 & 69.09 & 36.2 & | & | & | & LM & 2.0 \\ 
    M81X083 & 2CXO J095526.9+690541 & 148.863 & 69.095 & 36.5 & | & | & | & LM & 3.0 \\ 
    M81X084 & 2CXO J095527.2+690247 & 148.864 & 69.047 & 37.3 & | & | & | & LM & 2.0 \\ 
    M81X085 & 2CXO J095527.2+690250 & 148.864 & 69.047 & 36.3 & | & | & | & LM & 3.0 \\ 
    M81X086 & 2CXO J095527.5+690631 & 148.865 & 69.109 & 36.0 & -1.982 & | & 2.010 & LM & 2.0 \\ 
    M81X087 & 2CXO J095527.7+690704 & 148.866 & 69.118 & 36.3 & -3.491 & -0.087 & 0.323 & IM & 1.0 \\ 
    M81X088 & 2CXO J095527.7+690400 & 148.866 & 69.067 & 37.1 & | & | & | & LM & 2.0 \\ 
    M81X089 & 2CXO J095528.2+690541 & 148.867 & 69.095 & 35.9 & | & | & | & LM & 2.0 \\ 
    M81X090 & 2CXO J095528.4+690244 & 148.869 & 69.046 & 36.7 & | & | & | & (SNR) & 2.0 \\ 
    M81X091 & 2CXO J095528.8+690613 & 148.87 & 69.104 & 37.2 & -2.861 & | & 0.422 & LM & 2.0 \\ 
    M81X092 & 2CXO J095529.1+690320 & 148.871 & 69.056 & 36.8 & | & | & | & LM & 2.0 \\ 
    M81X093 & | & 148.873 & 69.047 & 35.8 & | & | & | & LM & 2.0 \\ 
    M81X094 & 2CXO J095530.1+690318 & 148.876 & 69.055 & 36.9 & | & | & | & LM & 2.0 \\ 
    M81X095 & 2CXO J095530.2+690246 & 148.876 & 69.046 & 36.5 & | & | & | & LM & 2.0 \\ 
    M81X096 & 2CXO J095530.3+690039 & 148.877 & 69.011 & 35.9 & -2.849 & | & 1.372 & IM & 1.0 \\ 
    M81X097 & | & 148.877 & 69.071 & 36.5 & | & | & | & LM & 2.0 \\ 
    M81X098 & | & 148.878 & 69.062 & 35.9 & | & | & | & LM & 2.0 \\ 
    M81X099 & | & 148.878 & 69.062 & 36.2 & | & | & | & LM & 2.0 \\ 
    M81X100 & 2CXO J095531.0+690055 & 148.879 & 69.015 & 35.8 & -4.512 & 0.092 & 0.006 & (SNR) & 1.0 \\ 
    M81X101 & 2CXO J095531.1+690144 & 148.88 & 69.029 & 35.9 & -4.377 & | & 1.695 & Gal & 1.0 \\ 
    M81X102 & 2CXO J095531.2+690418 & 148.881 & 69.072 & 37.6 & | & | & | & LM & 2.0 \\ 
    M81X103 & | & 148.885 & 69.069 & 36.3 & | & | & | & LM & 2.0 \\ 
    M81X104 & 2CXO J095532.6+690231 & 148.886 & 69.042 & 37.1 & -3.244 & | & 2.022 & IM & 1.0 \\ 
    M81X105 & 2CXO J095532.6+690352 & 148.886 & 69.065 & 37.8 & | & | & | & LM & 2.0 \\ 
    M81X106 & 2CXO J095532.6+690513 & 148.886 & 69.087 & 36.3 & | & | & | & (SNR) & 2.0 \\ 
    M81X107 & | & 148.886 & 69.073 & 36.1 & | & | & | & LM & 2.0 \\ 
    M81X108 & 2CXO J095532.8+690639 & 148.887 & 69.111 & 35.9 & -9.241 & 1.173 & 0.874 & \textit{LM} & 1.0 \\ 
    M81X110 & 2CXO J095532.9+690033 & 148.887 & 69.009 & 39.2 & -4.326 & 0.066 & -0.025 & IM & 1.0 \\ 
    M81X111 & | & 148.888 & 69.067 & 36.9 & | & | & | & LM & 2.0 \\ 
    M81X112 & | & 148.888 & 69.068 & 36.8 & | & | & | & LM & 2.0 \\ 
    M81X113 & 2CXO J095533.1+690354 & 148.888 & 69.065 & 39.4 & | & | & | & Nucleus & 1.0 \\ 
    M81X117 & 2CXO J095533.7+690124 & 148.891 & 69.023 & 36.3 & | & | & | & LM & 1.0 \\ 
    M81X118 & | & 148.892 & 69.065 & 36.8 & | & | & | & LM & 2.0 \\ 
    M81X119 & 2CXO J095534.0+690713 & 148.892 & 69.12 & 37.1 & -6.572 & 0.297 & 0.267 & HM & 1.0 \\ 
    M81X120 & 2CXO J095534.1+690618 & 148.892 & 69.105 & 35.9 & -3.622 & 0.695 & 0.146 & IM & 1.0 \\ 
    M81X121 & 2CXO J095534.3+690350 & 148.893 & 69.064 & 37.5 & | & | & | & LM & 2.0 \\ 
    M81X122 & | & 148.893 & 69.063 & 37.0 & | & | & | & LM & 2.0 \\ 
    M81X123 & | & 148.894 & 69.061 & 36.6 & -8.698 & 1.319 & 0.983 & \textit{LM} & 1.0 \\ 
    M81X124 & 2CXO J095534.5+690250 & 148.894 & 69.047 & 37.4 & | & | & | & LM & 2.0 \\ 
    M81X125 & 2CXO J095534.6+690453 & 148.894 & 69.082 & 37.5 & | & | & | & LM & 2.0 \\ 
    M81X126 & 2CXO J095534.7+690351 & 148.895 & 69.064 & 37.8 & | & | & | & LM & 2.0 \\ 
    M81X127 & 2CXO J095534.8+690408 & 148.895 & 69.069 & 37.5 & | & | & | & LM & 2.0 \\ 
    M81X128 & 2CXO J095534.9+690342 & 148.896 & 69.062 & 38.2 & | & | & | & LM & 2.0 \\ 
    M81X129 & 2CXO J095535.2+690316 & 148.897 & 69.054 & 37.8 & | & | & | & LM & 2.0 \\ 
    M81X130 & 2CXO J095535.3+690352 & 148.897 & 69.065 & 37.7 & | & | & | & LM & 2.0 \\ 
    M81X131 & 2CXO J095535.3+690638 & 148.898 & 69.111 & 37.2 & -4.819 & 0.88 & 1.699 & HM & 1.0 \\ 
    M81X132 & 2CXO J095535.4+690557 & 148.898 & 69.099 & 36.8 & | & | & | & LM & 2.0 \\ 
    M81X133 & 2CXO J095536.2+690245 & 148.901 & 69.046 & 36.5 & | & | & | & (SNR) & 2.0 \\ 
    M81X134 & 2CXO J095536.6+690632 & 148.903 & 69.109 & 37.2 & -7.988 & 1.351 & 1.010 & \textit{LM} & 1.0 \\ 
    M81X135 & 2CXO J095536.9+685656 & 148.904 & 68.949 & 36.9 & -2.325 & 0.508 & 1.195 & LM & 1.0 \\
    M81X136 & 2CXO J095536.9+690439 & 148.904 & 69.078 & 37.0 & | & | & | & LM & 2.0 \\ 
    M81X137 & 2CXO J095536.9+690433 & 148.904 & 69.076 & 37.5 & | & | & | & LM & 2.0 \\ 
    M81X138 & 2CXO J095537.2+690207 & 148.905 & 69.035 & 36.1 & -9.593 & 1.304 & 1.006 & \textit{LM} & 1.0 \\ 
    M81X139 & 2CXO J095537.5+690457 & 148.907 & 69.083 & 36.9 & | & | & | & LM & 2.0 \\ 
    M81X140 & 2CXO J095537.6+685833 & 148.907 & 68.976 & 36.6 & -4.106 & 0.853 & 0.945 & HM & 1.0 \\
    M81X141 & 2CXO J095537.7+690327 & 148.907 & 69.058 & 36.1 & -9.272 & 1.016 & 0.758 & \textit{LM} & 1.0 \\ 
    M81X142 & | & 148.91 & 68.943 & 36.5 & -2.544 & | & 1.228 & LM & 2.0 \\ 
    M81X143 & 2CXO J095538.9+690423 & 148.912 & 69.073 & 36.4 & | & | & | & LM & 2.0 \\ 
    M81X144 & 2CXO J095539.9+690348 & 148.916 & 69.064 & 36.5 & | & | & | & LM & 2.0 \\ 
    M81X145 & 2CXO J095540.3+690314 & 148.918 & 69.054 & 36.4 & | & | & | & LM & 2.0 \\ 
    M81X146 & 2CXO J095540.6+690105 & 148.92 & 69.018 & 36.1 & | & | & | & LM & 3.0 \\ 
    M81X147 & 2CXO J095540.7+690258 & 148.92 & 69.05 & 36.1 & | & | & | & LM & 2.0 \\ 
    M81X148 & 2CXO J095541.8+690301 & 148.925 & 69.051 & 35.9 & | & | & | & LM & 2.0 \\ 
    M81X149 & 2CXO J095541.9+690504 & 148.925 & 69.085 & 36.2 & | & | & | & LM & 3.0 \\ 
    M81X150 & 2CXO J095542.1+690336 & 148.926 & 69.06 & 38.0 & -6.606 & 0.307 & 0.364 & HM & 1.0 \\ 
    M81X151 & | & 148.926 & 69.116 & 36.2 & -2.067 & | & 1.567 & SNR & 1.0 \\ 
    M81X152 & 2CXO J095542.5+690320 & 148.927 & 69.056 & 37.1 & | & | & | & LM & 2.0 \\ 
    M81X153 & 2CXO J095542.5+691127 & 148.928 & 69.191 & 36.8 & -4.71 & 1.782 & 0.899 & (SNR) & 1.0 \\ 
    M81X154 & 2CXO J095542.9+690522 & 148.929 & 69.09 & 36.0 & | & | & | & LM & 1.0 \\ 
    M81X155 & 2CXO J095543.1+690445 & 148.93 & 69.079 & 36.6 & | & | & | & LM & 2.0 \\ 
    M81X156 & 2CXO J095543.2+690423 & 148.93 & 69.073 & 36.5 & | & | & | & LM & 2.0 \\ 
    M81X157 & 2CXO J095543.5+690355 & 148.932 & 69.065 & 36.3 & | & | & | & LM & 2.0 \\ 
    M81X158 & 2CXO J095543.8+690551 & 148.932 & 69.098 & 36.5 & -4.485 & 2.117 & 1.364 & IM & 1.0 \\ 
    M81X159 & 2CXO J095543.7+685905 & 148.932 & 68.985 & 37.4 & | & | & | & LM & 1.0 \\ 
    M81X160 & 2CXO J095544.5+690534 & 148.936 & 69.093 & 36.8 & -5.657 & 0.723 & 1.113 & HM & 1.0 \\
    M81X161 & 2CXO J095544.6+691003 & 148.936 & 69.168 & 36.9 & -6.616 & 0.128 & 0.049 & HM & 1.0 \\
    M81X162 & 2CXO J095545.3+690253 & 148.939 & 69.048 & 36.1 & -4.253 & 0.58 & 0.838 & (SNR) & 3.0 \\ 
    M81X163 & | & 148.939 & 69.038 & 35.9 & | & | & | & Gal & 1.0 \\ 
    M81X164 & 2CXO J095545.8+690300 & 148.941 & 69.05 & 37.4 & -10.078 & 1.154 & 0.829 & \textit{LM} & 1.0 \\ 
    M81X165 & 2CXO J095547.0+690551 & 148.946 & 69.098 & 37.9 & -9.11 & 1.519 & 1.164 & \textit{LM} & 1.0 \\ 
    M81X166 & 2CXO J095546.9+690536 & 148.946 & 69.093 & 36.3 & -3.238 & | & 1.865 & IM & 1.0 \\ 
    M81X167 & 2CXO J095547.9+685928 & 148.95 & 68.991 & 36.3 & -5.454 & 0.433 & 0.016 & SNR & 1.0 \\
    M81X168 & 2CXO J095548.2+685915 & 148.951 & 68.988 & 36.4 & -3.217 & | & 1.696 & IM & 2.0 \\ 
    M81X169 & 2CXO J095548.7+690140 & 148.953 & 69.028 & 36.2 & -2.954 & | & 2.318 & IM & 2.0 \\ 
    M81X170 & 2CXO J095549.3+685836 & 148.956 & 68.977 & 37.6 & -6.873 & 0.694 & 0.438 & HM & 1.0 \\
    M81X171 & 2CXO J095549.4+690811 & 148.956 & 69.137 & 38.0 & -2.223 & | & 1.808 & LM & 1.0 \\ 
    M81X172 & 2CXO J095549.7+690531 & 148.958 & 69.092 & 38.7 & -7.053 & 1.218 & 0.915 & \textit{LM} & 1.0 \\ 
    M81X173 & 2CXO J095549.8+690300 & 148.958 & 69.05 & 36.1 & -3.744 & | & 0.797 & IM & 2.0 \\ 
    M81X174 & 2CXO J095550.0+690714 & 148.959 & 69.121 & 36.5 & -5.162 & 0.162 & 0.104 & HM & 2.0 \\
    M81X175 & 2CXO J095550.1+690540 & 148.959 & 69.095 & 36.3 & | & | & | & LM & 1.0 \\ 
    M81X176 & 2CXO J095550.5+685832 & 148.961 & 68.976 & 36.7 & -2.688 & | & 1.234 & LM & 1.0 \\ 
    M81X177 & 2CXO J095551.0+690512 & 148.963 & 69.087 & 36.4 & | & | & | & (SNR) & 2.0 \\ 
    M81X178 & 2CXO J095551.5+691104 & 148.964 & 69.185 & 36.7 & -2.902 & 2.969 & 1.241 & IM & 1.0 \\
    M81X179 & 2CXO J095551.5+685910 & 148.965 & 68.986 & 36.0 & -5.353 & 0.356 & 0.416 & HM & 2.0 \\
    M81X180 & 2CXO J095551.8+690739 & 148.966 & 69.128 & 36.7 & -9.302 & 1.285 & 0.951 & \textit{LM} & 1.0 \\ 
    M81X181 & 2CXO J095552.4+685625 & 148.968 & 68.94 & 36.5 & -2.508 & -2.176 & 2.398 & IM & 1.0 \\
    M81X182 & 2CXO J095552.4+690306 & 148.969 & 69.052 & 38.2 & | & | & | & LM & 2.0 \\ 
    M81X183 & 2CXO J095553.1+685926 & 148.971 & 68.991 & 37.8 & -3.545 & -0.28 & -0.302 & HM & 1.0 \\ 
    M81X184 & 2CXO J095553.2+690207 & 148.972 & 69.035 & 36.9 & -5.731 & 1.087 & 1.019 & HM & 1.0 \\
    M81X185 & 2CXO J095553.3+690446 & 148.972 & 69.08 & 36.1 & -4.86 & 1.256 & 0.993 & (SNR) & 1.0 \\ 
    M81X186 & 2CXO J095553.6+690434 & 148.973 & 69.076 & 37.1 & | & | & | & LM & 1.0 \\ 
    M81X187 & 2CXO J095554.2+690346 & 148.976 & 69.063 & 36.3 & | & | & | & (SNR) & 1.0 \\ 
    M81X188 & 2CXO J095554.9+690055 & 148.979 & 69.016 & 37.5 & -9.38 & 1.376 & 1.098 & \textit{LM} & 1.0 \\ 
    M81X189 & 2CXO J095554.9+690239 & 148.979 & 69.044 & 36.9 & -5.432 & 1.408 & 1.348 & Gal & 2.0 \\ 
    M81X190 & 2CXO J095555.5+691007 & 148.981 & 69.169 & 36.7 & -4.442 & 0.875 & 0.815 & HM & 1.0 \\
    M81X191 & 2CXO J095555.3+685859 & 148.981 & 68.983 & 36.1 & -4.784 & 0.355 & 0.353 & (SNR) & 2.0 \\ 
    M81X192 & 2CXO J095555.6+690814 & 148.982 & 69.137 & 36.9 & -2.299 & | & 1.085 & LM & 1.0 \\ 
    M81X193 & 2CXO J095555.7+690901 & 148.982 & 69.15 & 36.4 & -3.554 & 0.545 & 0.422 & IM & 1.0 \\ 
    M81X194 & 2CXO J095555.9+690515 & 148.983 & 69.088 & 36.3 & -2.578 & | & 1.946 & LM & 2.0 \\ 
    M81X195 & 2CXO J095556.0+690358 & 148.984 & 69.066 & 36.1 & | & | & | & (SNR) & 2.0 \\ 
    M81X196 & 2CXO J095556.0+690312 & 148.984 & 69.053 & 36.6 & | & | & | & (SNR) & 1.0 \\ 
    M81X197 & 2CXO J095556.5+690802 & 148.986 & 69.134 & 37.8 & -5.53 & 1.17 & 1.417 & \textit{LM} & 2.0 \\ 
    M81X198 & 2CXO J095557.6+690436 & 148.99 & 69.077 & 37.8 & -4.561 & 0.024 & | & IM & 1.0 \\ 
    M81X199 & 2CXO J095558.5+690525 & 148.994 & 69.091 & 37.6 & | & | & | & LM & 1.0 \\ 
    M81X200 & 2CXO J095559.1+690617 & 148.997 & 69.105 & 37.2 & -3.48 & | & 1.190 & IM & 1.0 \\ 
    M81X201 & | & 149.0 & 69.093 & 36.2 & -6.115 & 2.352 & 1.376 & Gal & 1.0 \\ 
    M81X202 & 2CXO J095600.1+690418 & 149.001 & 69.072 & 36.5 & -3.374 & -0.105 & | & (SNR) & 1.0 \\ 
    M81X203 & | & 149.001 & 69.021 & 35.6 & -4.979 & 2.101 & 1.158 & HM & 2.0 \\ 
    M81X204 & | & 149.007 & 69.005 & 35.8 & -5.234 & -0.184 & 0.055 & HM & 1.0 \\ 
    M81X205 & 2CXO J095601.9+685859 & 149.008 & 68.983 & 37.0 & -6.296 & 0.283 & 0.324 & HM & 1.0 \\
    M81X206 & 2CXO J095602.6+685935 & 149.011 & 68.993 & 37.6 & -4.437 & 0.137 & 0.810 & HM & 2.0 \\
    M81X207 & 2CXO J095602.7+685844 & 149.011 & 68.979 & 37.1 & -5.699 & 1.976 & -0.487 & HM & 1.0 \\
    M81X208 & 2CXO J095602.6+690547 & 149.011 & 69.096 & 36.6 & -2.912 & | & 1.610 & IM & 1.0 \\ 
    M81X209 & 2CXO J095602.9+690217 & 149.013 & 69.038 & 36.7 & -3.485 & 1.138 & 1.664 & IM & 2.0 \\
    M81X210 & 2CXO J095603.2+690107 & 149.013 & 69.019 & 36.4 & -1.603 & | & 2.348 & LM & 1.0 \\ 
    M81X211 & | & 149.014 & 69.001 & 36.2 & -6.076 & 0.629 & 0.506 & HM & 1.0 \\ 
    M81X212 & 2CXO J095604.0+690726 & 149.017 & 69.124 & 36.3 & -2.56 & | & 2.097 & LM & 1.0 \\ 
    M81X213 & 2CXO J095604.8+690344 & 149.02 & 69.062 & 36.3 & -8.314 & 0.044 & 0.120 & (SNR) & 1.0 \\ 
    M81X214 & 2CXO J095604.7+685840 & 149.02 & 68.978 & 36.9 & -3.822 & 3.515 & 1.616 & IM & 1.0 \\ 
    M81X215 & | & 149.023 & 69.112 & 36.2 & -8.904 & 1.192 & 0.918 & \textit{LM} & 1.0 \\
    M81X216 & 2CXO J095606.0+685941 & 149.025 & 68.995 & 37.0 & -4.358 & 1.851 & 1.720 & Gal & 1.0 \\
    M81X217 & 2CXO J095606.0+690833 & 149.025 & 69.143 & 36.8 & -5.335 & 0.712 & 0.606 & HM & 1.0 \\
    M81X218 & 2CXO J095607.7+690325 & 149.032 & 69.057 & 37.0 & -4.637 & 0.74 & 1.172 & HM & 1.0 \\ 
    M81X219 & 2CXO J095608.1+690142 & 149.034 & 69.029 & 36.0 & -3.381 & 1.961 & 1.163 & IM & 1.0 \\
    M81X220 & 2CXO J095608.9+690106 & 149.037 & 69.019 & 37.6 & -5.432 & -0.141 & 0.149 & HM & 1.0 \\ 
    M81X221 & | & 149.038 & 68.993 & 36.0 & -3.922 & -0.058 & -0.181 & HM & 2.0 \\ 
    M81X222 & | & 149.052 & 69.104 & 36.4 & -2.258 & | & 1.823 & LM & 2.0 \\ 
    M81X223 & 2CXO J095613.6+690631 & 149.056 & 69.109 & 37.3 & -2.259 & 0.716 & 1.094 & LM & 1.0 \\
    M81X224 & 2CXO J095613.7+685724 & 149.057 & 68.957 & 37.4 & | & | & | & Star & 1.0 \\ 
    M81X225 & 2CXO J095614.3+690248 & 149.06 & 69.047 & 37.1 & | & | & | & LM & 1.0 \\ 
    M81X226 & 2CXO J095614.6+690339 & 149.061 & 69.061 & 36.8 & -4.918 & 1.272 & 1.047 & Gal & 1.0 \\ 
    M81X227 & 2CXO J095614.9+685732 & 149.062 & 68.959 & 37.3 & -5.357 & 0.913 & 0.523 & HM & 1.0 \\
    M81X228 & 2CXO J095616.3+690119 & 149.068 & 69.022 & 36.4 & -2.977 & | & 1.805 & IM & 1.0 \\ 
    M81X229 & 2CXO J095616.5+685649 & 149.069 & 68.947 & 37.1 & -3.034 & 0.886 & 0.936 & IM & 1.0 \\
    M81X230 & 2CXO J095617.0+685820 & 149.071 & 68.972 & 37.5 & -5.536 & 0.901 & 0.374 & HM & 1.0 \\
    M81X231 & | & 149.076 & 69.052 & 36.8 & -2.407 & | & 2.423 & LM & 1.0 \\ 
    M81X232 & 2CXO J095619.7+690201 & 149.083 & 69.034 & 36.9 & -3.746 & 0.79 & 0.482 & IM & 1.0 \\ 
    M81X233 & 2CXO J095622.2+690220 & 149.092 & 69.039 & 36.5 & -3.226 & | & 1.395 & IM & 1.0 \\ 
    M81X234 & 2CXO J095622.3+690446 & 149.093 & 69.08 & 36.6 & -2.734 & | & 1.652 & (SNR) & 2.0 \\ 
    M81X235 & | & 149.098 & 69.03 & 36.1 & -2.855 & 1.09 & 2.473 & IM & 1.0 \\ 
    M81X236 & 2CXO J095624.0+690009 & 149.1 & 69.003 & 36.6 & -4.584 & 0.512 & 0.778 & HM & 1.0 \\ 
    M81X237 & 2CXO J095628.0+690102 & 149.117 & 69.017 & 37.4 & -5.536 & 0.273 & 0.599 & HM & 1.0 \\
    M81X238 & 2CXO J095630.8+690222 & 149.128 & 69.04 & 36.8 & -2.48 & | & 1.762 & LM & 1.0 \\ 
    M81X239 & 2CXO J095631.0+685837 & 149.129 & 68.977 & 37.0 & -2.224 & 1.037 & 2.038 & LM & 1.0 \\
    M81X240 & 2CXO J095632.5+685714 & 149.135 & 68.954 & 36.9 & | & | & | & LM & 1.0 \\ 
    M81X241 & 2CXO J095633.4+690035 & 149.139 & 69.01 & 36.5 & -2.098 & | & 1.464 & LM & 2.0 \\ 
    M81X242 & 2CXO J095633.5+690331 & 149.14 & 69.059 & 37.3 & -1.704 & 1.187 & 1.972 & LM & 1.0 \\ 
    M81X243 & 2CXO J095636.4+690028 & 149.152 & 69.008 & 37.9 & -8.657 & 1.029 & 0.779 & HM & 1.0 \\
    M81X244 & 2CXO J095642.0+685858 & 149.174 & 68.983 & 36.7 & -4.448 & -0.224 & -0.081 & HM & 1.0 \\ 
    M81X245 & 2CXO J095645.2+690108 & 149.189 & 69.019 & 36.7 & -2.038 & -0.169 & 1.207 & LM & 1.0 \\ 
    M81X246 & | & 149.198 & 69.064 & 36.9 & -2.889 & 1.17 & 0.746 & IM & 1.0 \\ 
    M81X248 & | & 149.214 & 68.977 & 36.6 & -4.075 & 0.378 & 0.588 & IM & 1.0 \\ 
    M81X249 & 2CXO J095651.4+685606 & 149.214 & 68.935 & 37.8 & -2.469 & | & 1.652 & LM & 1.0 \\ 
    M81X250 & | & 149.227 & 69.032 & 36.9 & -2.051 & | & 1.344 & LM & 1.0 \\ 
    M81X251 & 2CXO J095658.2+690046 & 149.243 & 69.013 & 37.5 & -7.659 & 0.517 & 0.760 & HM & 1.0 \\
\enddata
\tablecomments{Properties and classifications of all M81 X-ray sources identified in \citet{lehmer19} that fall within the footprint of the HST image. For each source, the ID is the galaxy name and the ID number assigned in \citet{lehmer19}. X-ray luminosities are in units log \es, and magnitudes are absolute mags estimated at a distance of 3.63 Mpc. The classification for each source is given, with (SNR) representing SNRs identified using our HR-L$_{\rm{X}}$ criterion, and italics indicating XRBs associated with clusters. A confidence flag (CF) is assigned to each source based on the ``strength" of the identification of the X-ray emitter: a CF of 1 represents the most certain classifications (i.e. those determined in other studies, or XRBs with a clear donor, with multiple candidates of similar mass, or a clear absence of a donor); CF ratings of 2 or 3 may indicate that a source is in a dust-obscured region, such as near the nucleus or along a dust lane, since the presence of heavy dust could potentially mask high-mass stars, background galaxies, and clusters, leading to possible misidentifications.} 
\end{deluxetable*}

%% file: mosaics.tex
\begin{figure*}
\includegraphics[width=\linewidth]{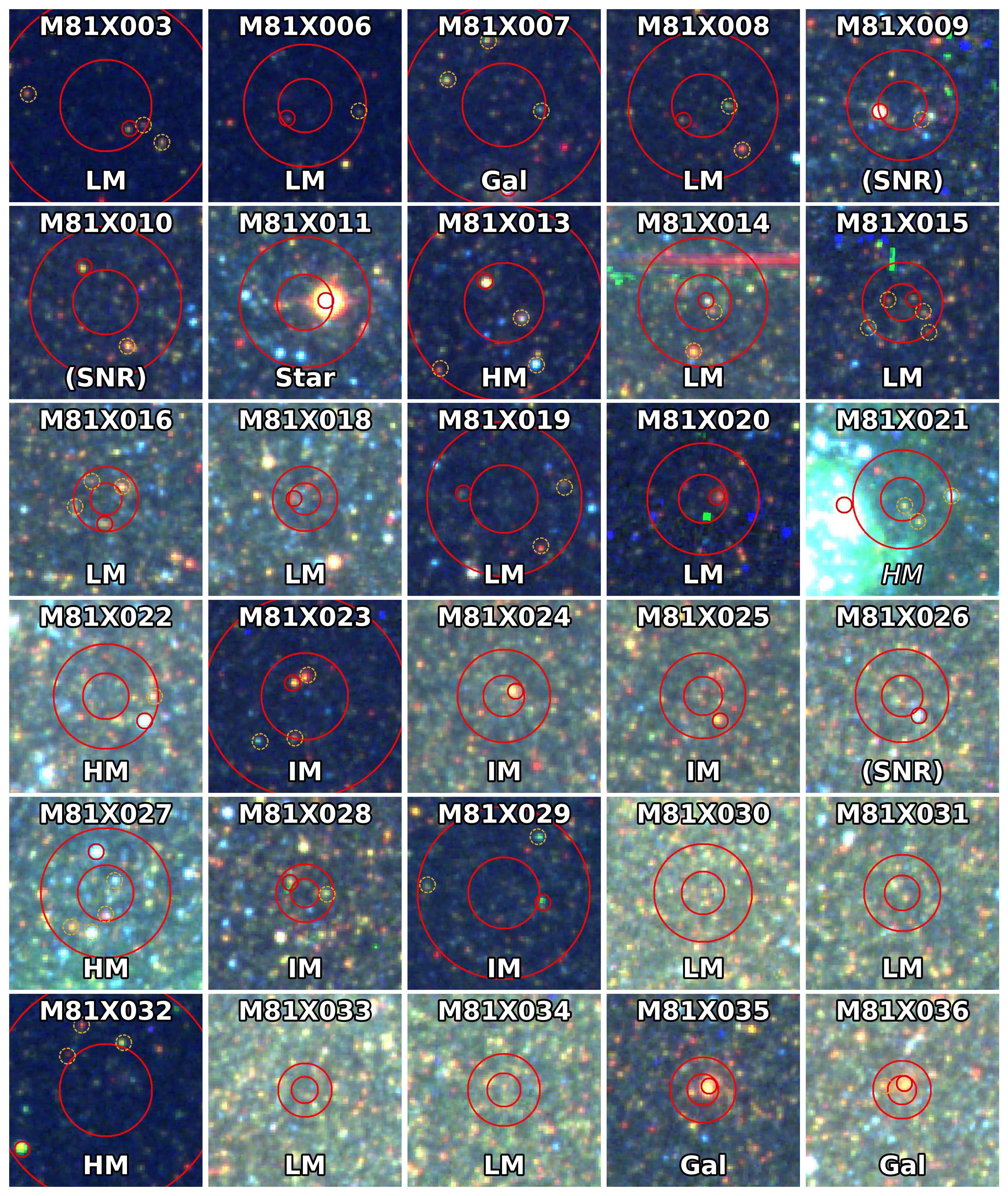}
\caption{Identified point sources within the 1- and 2-$\sigma$ radii for each X-ray source, with the most likely donor circled in red. The classifications of each source are also given, with LM, IM, and HM representing our \lms, \ims, and \hms, respectively, and those in italics represent cluster XRBs. Sources classified as (SNR) represent SNRs identified using our HR-L$_{\rm{X}}$ criterion; all other SNRs are identified within published catalogs \citep[][see \S\ref{sec:snr}]{nantais10,nantais11,sc10}. The size of each image is roughly $3.7\times3.7$ arcseconds.\label{fig:mosaics}}
\end{figure*}

\begin{figure*}
\ContinuedFloat
\includegraphics[width=\linewidth]{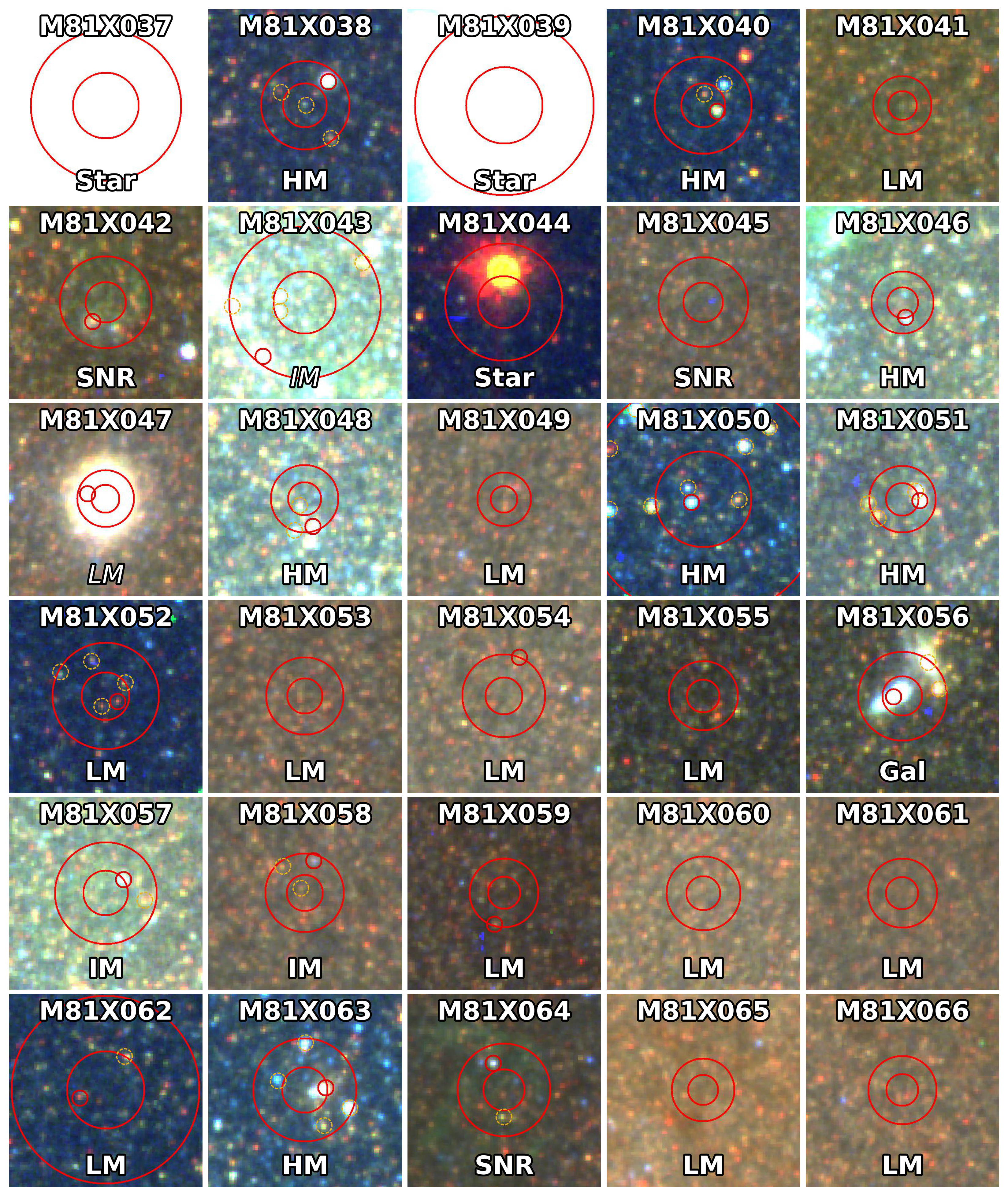}
\caption{\textit{(continued)}}
\end{figure*}

  \begin{figure*}
\ContinuedFloat
\includegraphics[width=\linewidth]{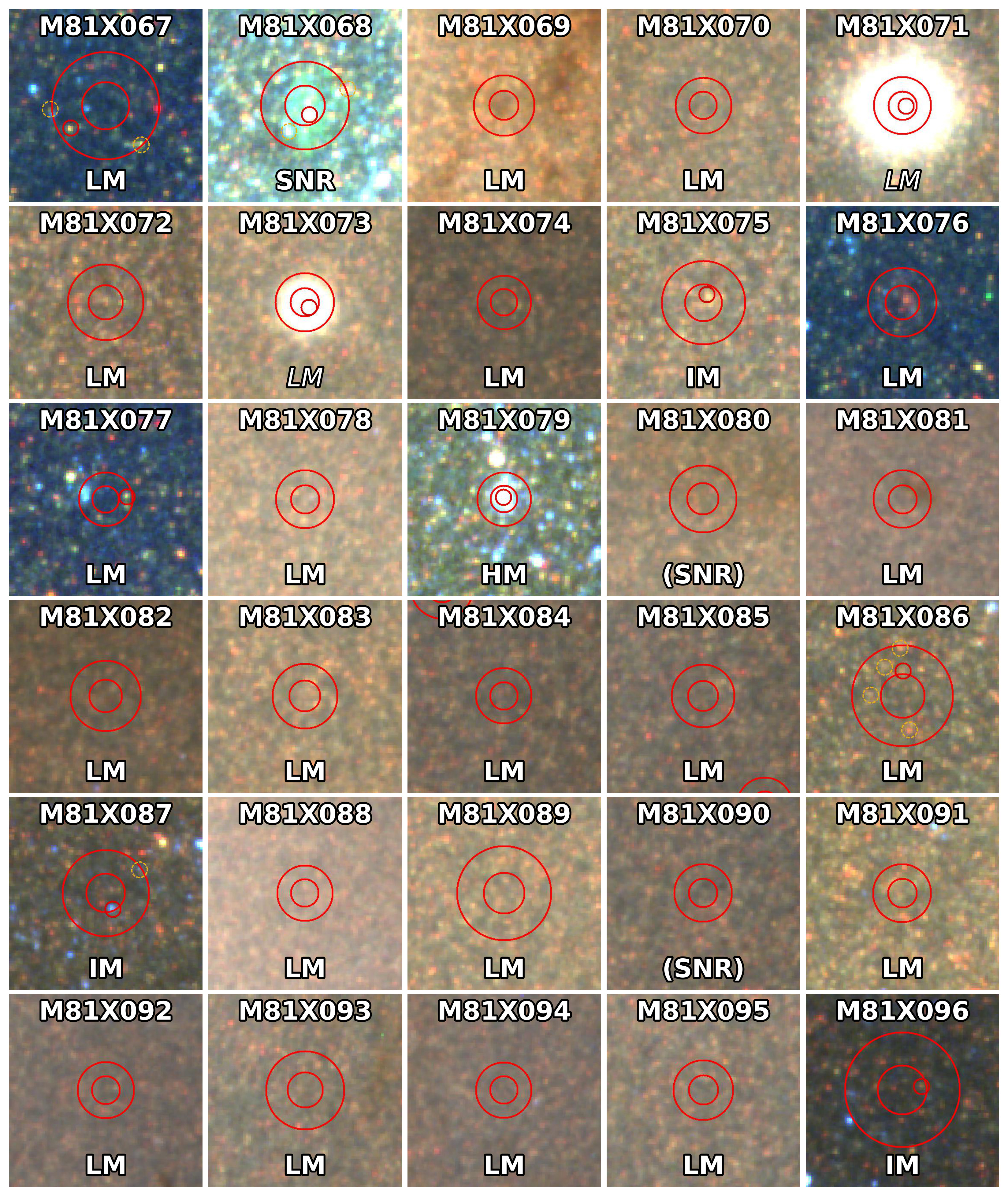}
\caption{\textit{(continued)}}
\end{figure*}

  \begin{figure*}
\ContinuedFloat
\includegraphics[width=\linewidth]{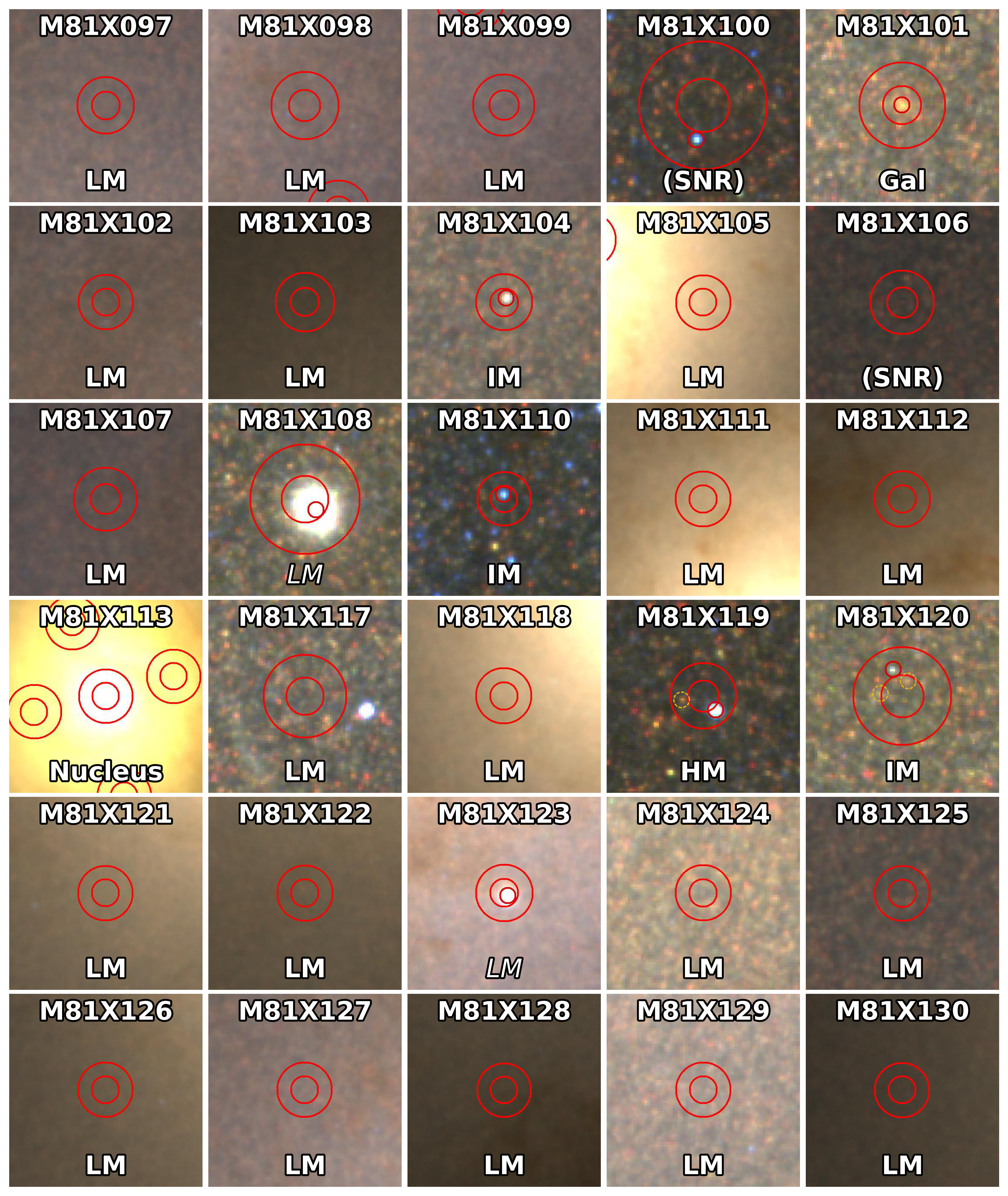}
\caption{\textit{(continued)}}
\end{figure*}

   \begin{figure*}
 \ContinuedFloat
\includegraphics[width=\linewidth]{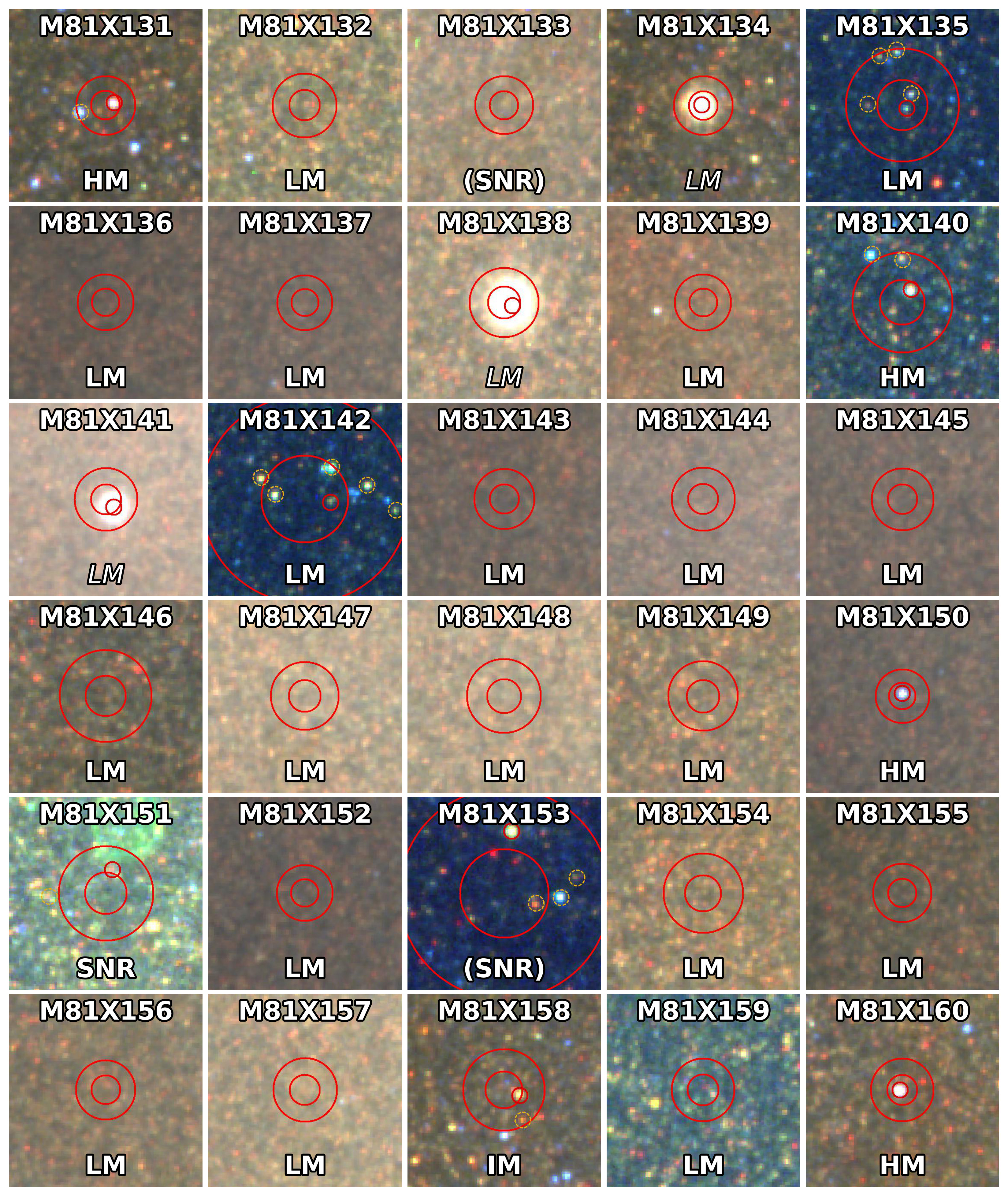}
\caption{\textit{(continued)}}
\end{figure*} 

  \begin{figure*}
\ContinuedFloat
\includegraphics[width=\linewidth]{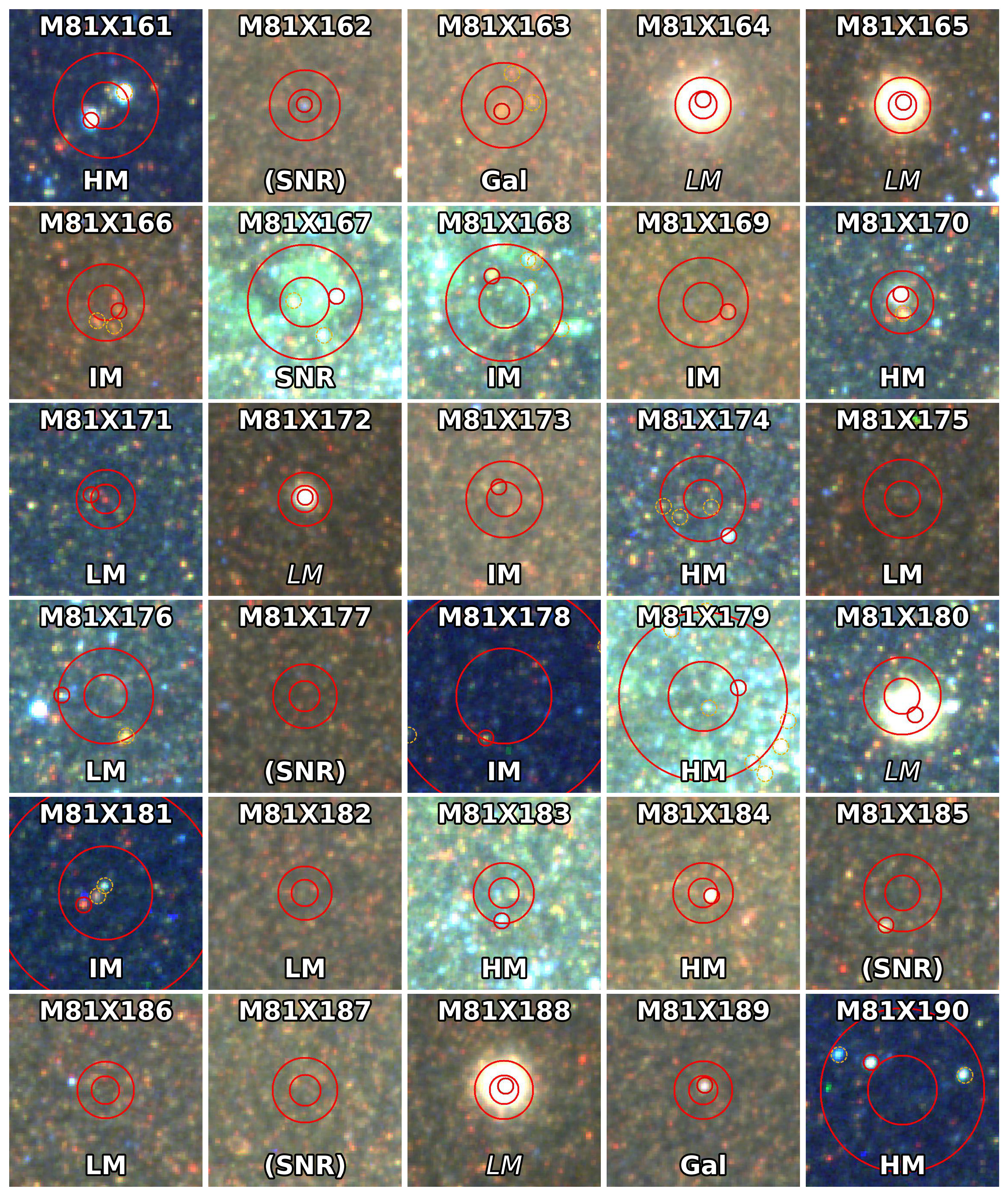}
\caption{\textit{(continued)}}
\end{figure*}   

  \begin{figure*}
\ContinuedFloat
\includegraphics[width=\linewidth]{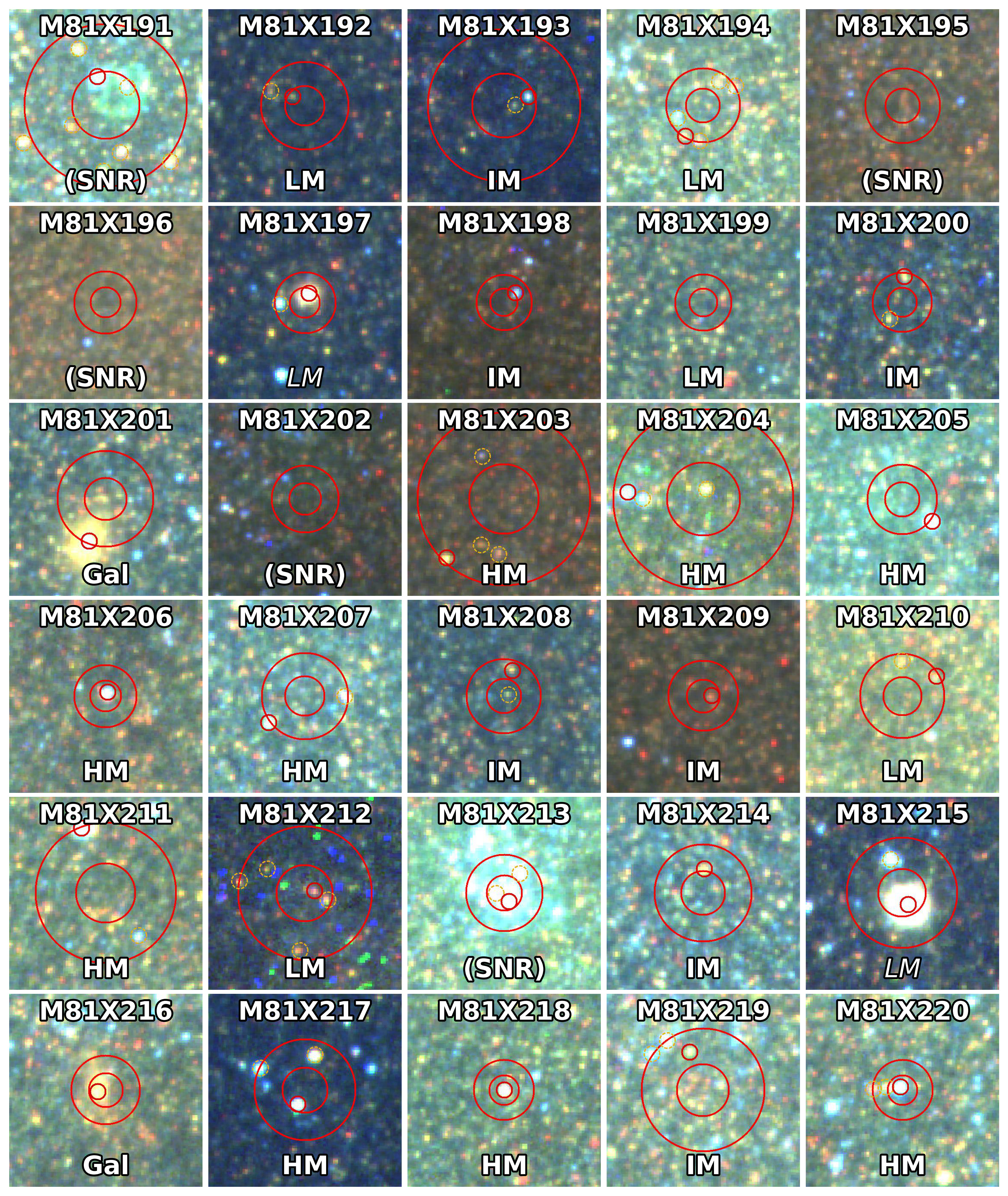}
\caption{\textit{(continued)}}
\end{figure*}

  \begin{figure*}
\ContinuedFloat
\includegraphics[width=\linewidth]{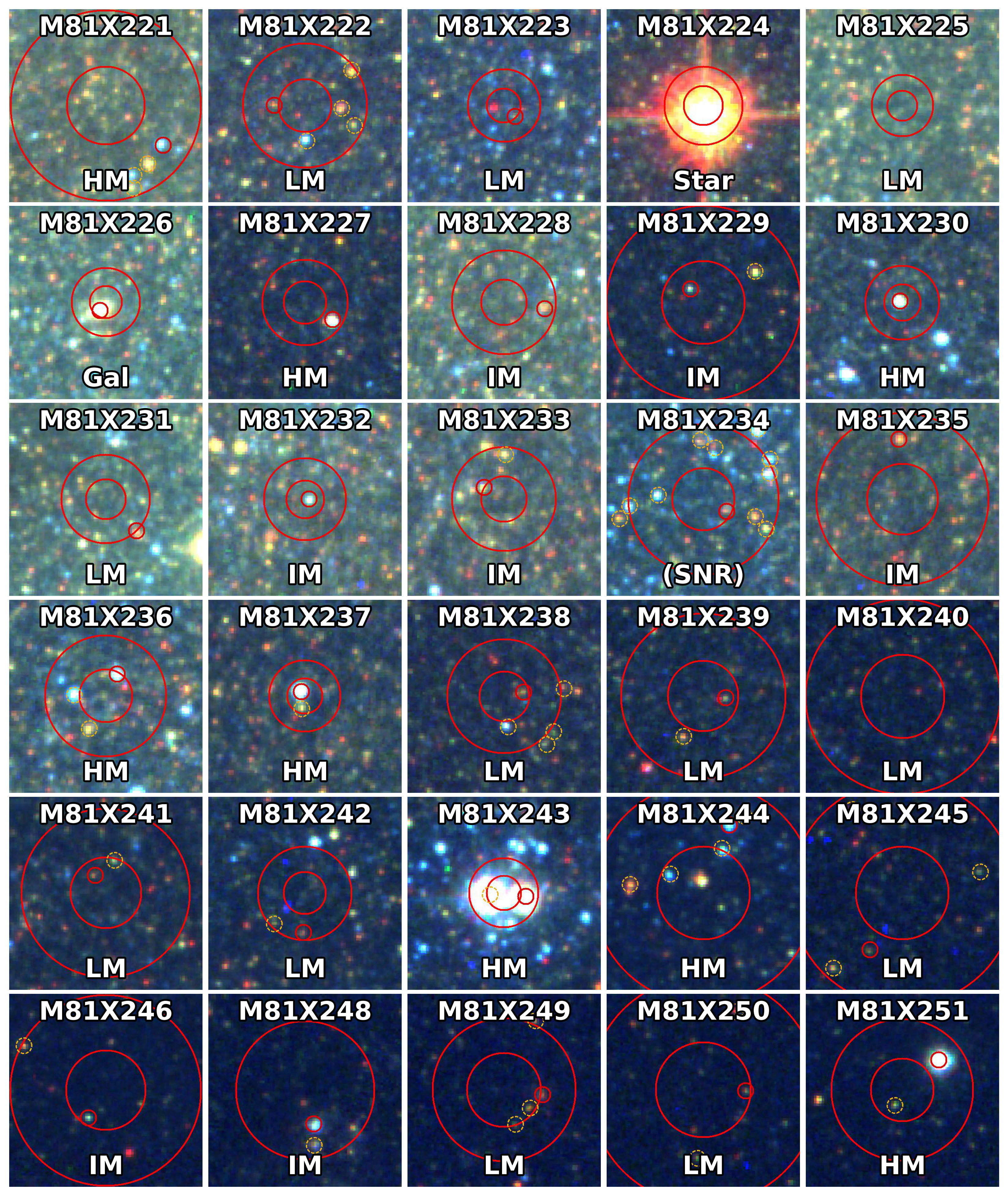}
\caption{\textit{(continued)}}
\end{figure*}